\documentclass[preprint,12pt]{elsarticle}

\usepackage[left=2.0cm, right=2.0cm, top=2.0cm, bottom=2.0cm]{geometry}
\usepackage{amssymb}
\usepackage{booktabs,latexsym}
\usepackage{graphicx}
\graphicspath{ {images/} }
\usepackage{amsmath}
\usepackage{multirow}
\usepackage{color}
\usepackage{array,booktabs}
\usepackage{float}
\usepackage{tabularx}
\usepackage{tabu}
\usepackage{caption}
\usepackage{subcaption}

\usepackage{arydshln}
\usepackage[latin1]{inputenc}
\usepackage{tikz}
\usetikzlibrary{shapes,arrows}

\newcommand{\s}[2]{$#1\times 10^{#2}$}

\newcommand*\rfrac[2]{{}^{#1}\!/_{#2}}
\newcommand{\ofp}{(\phi)}

\tikzstyle{empty} = []
\tikzstyle{ic} = [rectangle, draw, 
minimum width=6cm, text centered, rounded corners, minimum height=1cm]
\tikzstyle{solution} = [ellipse, draw, 
minimum width=2cm, text centered, rounded corners, minimum height=1cm]
\tikzstyle{solver} = [draw, rectangle, node distance=2cm,
minimum height=1cm, minimum width = 5cm]
\tikzstyle{line} = [draw, -latex']

\begin{document}
	\begin{frontmatter}
       \title{A mass-preserving level set method for simulating 2D/3D fluid flows with evolving interface}
\author[a]{H. L. Wen}
\author[b]{C. H. Yu}
\author[a,c,d]{Tony W. H. Sheu\corref{author1}}

\cortext[author1] {Corresponding author.\\\textit{E-mail address:} twhsheu@ntu.edu.tw}
\address[a]{Department of Engineering Science and Ocean Engineering, National Taiwan University, Taipei, R.O. China}
\address[b]{State key lab of Hydraulics and Mountain River Engineering, Sichuan University, Sichuan 610000, P.R. China}
\address[c]{Institute of Applied Mathematical Sciences, National Taiwan University, Taipei, R. O. China}
\address[d]{Center for Advanced Study in Theoretical Sciences, National Taiwan University, Taipei, R.O. China}
       \begin{abstract}
       	Within the context of Eulerian approaches, we aim to develop a new interface-capturing solver to predict two-phase flow in 2D/3D Cartesian meshes. To achieve mass conservation and to capture interface topology accurately, a mass-preserving level set advection equation cast in the scalar sign-distance function is developed. The novelty of the proposed Eulerian solver lies in the introduction of a scalar speed function to rigorously reconstruct the classical level set equation. Through several benchmark problems, the proposed flow solver for solving incompressible two-phase viscous flow equations has been verified.
       \end{abstract}
	\end{frontmatter}
  \section{Introduction}
  Two-phase flow is a field of fluid mechanics that has been extensively studied in the past because of its practical importance and computational challenge. In industry, enormous applications have been known to involve moving interfaces in fluids with different phases, liquid and gas phases for example. \\
  
  Remarkable developments have been made with regard to modeling formulations, namely the Lagrangian and Eulerian classes of numerical approaches. Lagrangian methods, such as front-tracking method \cite{Unverdi(1992),Glimm(1986)} and marker method \cite{Daly(1969)}, make use of markers to follow explicitly the interfaces (or fronts). Regridding algorithms of different sorts are normally required to prevent marker particles from clustering together. In addition, this class of methods is computationally rather expensive with the increasing number of particles. Moreover, management of addition or deletion of markers at a time when interface becomes largely stretched or deformed by the fluid flow is practically difficult and the simulation of this class of flows requires special treatment.\\
  
  Eulerian method is referred to as the other class of approaches to simulate two-phase flow problems. This class of methods is featured with the use of a scalar function to define the location of the interface. The level set methods due to Osher and Sethian in 1988 \cite{Osher(1988)}, Sussman et al. in 1994 \cite{Sussman(1994)} and Sethian in 1999 \cite{Sethian(1999)} employed smooth distance function with zero value of the level set function to denote the interface location. Level set method has a prevailing advantage in solving incompressible two-phase flow equations incorporating surface tension owing to its great ability to calculate both the curvature and surface normal of the interface more easily and accurately. However, the classical level set method suffers from poor mass conservation as Sussman et al. pointed out in the papers of \cite{Sussman(1994), Sussman(1998)}. How to preserve mass conservation and retain level set function as a distance function motivates the present development of a new advection algorithm for the transport of level set scalar function.\\
  
   There are several existing methods derived from the classical level set method to overcome the drawback of non-conservation of mass, for example, conservative level set method, purposed by Olsson et al. \cite{Olsson(2005)}, which works on the function being zero on one side of the interface and one on the other side. Interface-correction level set method, proposed by Zhouyang Ge et al. \cite{Ge(2018)}, solves an additional equation to correct the mass loss of level set function after solving the advection equation.\\
  
  Volume of Fluid (VOF) method of Hirt and Nichols \cite{Hirt(1981)}, Youngs \cite{Youngs(1982)}, Lafaurie et al. \cite{Lafaurie(1994)} is another popular and effective Eulerian method. The local volume of fraction of one of the fluids is chosen in flow domain to update the position of the interface. Across the interface, the volume fraction is sharply varied from 0 to 1 or vice versa. Since the interface is represented in terms of volume fraction, mass in principle is conserved all the time. The disadvantage of the VOF method is that it is more difficult to compute the local geometrical quantities at the interface from the volume fraction due the involvement of a sharp transition across the interface.\\
  
  While the level set method does not have the same ability of conserving mass as the VOF method, it has a good ability to accurately compute local surface normal and curvature and, therefore, it facilitators to capture topology change due to a sharp change in surface tension. As a result, to achieve mass conservation and capture interface accurately, the idea of combining the LS and VOF methods has been proposed to yield the well known CLSVOF method \cite{Bourlioux(1995),Sussman(2000)}. It is noted that in the CLSVOF method the level set function is used solely to compute the geometric properties at the interface, while the volume fraction is calculated from the VOF advection equation. The hybrid particle level set method, which is another hybrid method developed by Douglas Enright et al. \cite{Enright(2002)}, defines two sets of particles near the interface, and then detects the "escaped" particles to reconstruct the level set function. \\

It is noted that application of the hybrid method usually need more CPU time, even more than the sum of each method. For example, the total CPU time to solve the advection of LS method and the VOF method is usually less than the time used to solve of CLSOVF advection. Since the hybrid method needs an additional procedure to make sure that the two methods are consistent in the sense that the level set function and volume fraction function should have the same position of interface in CLSOVF method. Moreover, the hybrid method could be very complicated to implement it. As a result, we are motivated to develop an easy-to-implement mass-preserving level set model to simulate two phase flows, which does not require much modifications on classical level set method. The issue of retaining computational efficiency is also considered.\\

  The rest of this paper is organized as follows. In section 2, we present the interface capturing level set method that is applicable only to finite difference cells filled with the same fluid. Then, in section 2.2, we present a mass-preserving level set method that will be used solely in cells containing two different fluids, which are separated by an interface. The novelty of this newly proposed method presented in section 2.3 lies in the introduction of speed function to reconstruct the level set advection equation. In section 2.4, the re-initialization equation is described and it will be applied frequently to maintain the level set function to be a distance function. As a result, the interface can be more accurately captured even in cases when the interface has been largely stretched or deformed by the flow. In section 3, within the framework of one-field formulation in Cartesian coordinate system, the conservation equations for mass and momentum coupled with the mass-preserving level set equation described in section 2 shall be solved together. In section 4, the discretization schemes developed for our proposed two-phase flow solver are described. In section 5, the proposed two-phase flow solver will be justified to verify its accuracy through several benchmark problems. Section 6 describes two practical problems under current investigation. Conclusions are drawn in section 7.
  
  \section{Interface evolving equation} 
  Interface evolution is often the key of research in science and engineering contexts, such as two-phase flow. It is therefore essential to accurately model the evolution of the interface under a velocity field. One can model the moving interfaces more easily using the explicit techniques by solving a system of ordinary differential equations for the coordinate of a node $i$ as $d\mathbf{\underline{x}}^{(i)}/dt = \mathbf{\underline{v}}(\mathbf{\underline{x}^{(i)}},t)$, which will be sought subject to an initial condition $\mathbf{\underline{x}^{(i)}}(t=0)=\mathbf{\underline{x}_0^{(i)}}$, where $\mathbf{\underline{v}}$ is the velocity vector. This simple approach is, however, adequate only for a case with small deformations on the initial interface. Several drawbacks have been pointed out for some general flow motions \cite{Sethian(2001),Osher(2001)}.
  
 \subsection{Level set advection equation }
The implicit approach is another potential class of methods for choice to depict an evolving interface. The level set method is the most popular one  and will be adopted in this study. In the level set method, we implicitly represent the interface by the zero level set value of a smooth function $\phi(\mathbf{x})=0$ for all $\mathbf{x}$ in the flow domain $\Omega$. It is worthy to address here that in the level set method, a Cartesian grid is normally chosen to constitute the background mesh. It is also noted that the level set method has advantages of replacing the advection of physical properties with sharp gradients at the interface with the advection of level set function that is essentially smooth in nature. Within the context of implicitly representing the level set function $\phi$, its zero level set value is advected as follows by the velocity field $\mathbf{U}$
  
  \begin{equation}
  	\label{ls_advect}
  	\frac{\partial\phi}{\partial t} + \mathbf{U}\cdot\nabla\phi=0.
  \end{equation}
Note that all other level set values are advected by the same advection equation shown above. While the level set method has a good ability to compute the curvature of the interface more easily and accurately and is thus advantageous to be applied to capture topological change. This method does not necessarily achieve the same degree of conservation property as the VOF method or front tracking method. A model that can retain mass conservation in the case of involving an evolving interface should be employed with the level set method.\\
  
  \subsection{Mass-preserving level set advection equation}
  Within the level set method, a new mass-preserving level set advection algorithm for an implicit representation of the level set function $\phi$ will be presented below. In a cell containing interface, application of the classical level set equation solely can not ensure mass conservation due to sharp gradients established near the interface. Therefore, the level set value needs to be redistributed using Eq. (\ref{ls_advect}), in particular, in the vicinity of the interface through a proper addition of source term only to cells containing a line of interface (for two-dimensional problems) and a surface of interface (for three-dimensional problems).\\
  
Our underlying strategy of model development is to modify the interface slightly in the direction normal to the interface. To this end, a scalar speed function $\mathbb{F}_s$ is employed such that the evolution of interface is directed toward the direction normal to the interface itself. It is therefore meant that $\mathbf{U}$ in (\ref{ls_advect}) is identical to $\mathbf{U}=\mathbb{F}_s\hat{n}$. The normal direction $\hat{n}$ can be expressed in terms of $\phi$ as $\hat{n}= \rfrac{\nabla\phi}{|\nabla\phi|}$. This geometric variable is the result of the fact that $\phi$ is constant at a level set and $\nabla\phi$ points in the direction normal to the interface. By substituting $\mathbf{U}=\mathbb{F}_s\hat{n}=\mathbb{F}_s \rfrac{\nabla\phi}{|\nabla\phi|}$ into Eq. (\ref{ls_advect}), we can get the corresponding level set equation given below

  \begin{equation}
  	\label{ls_advect_fs}
  	\frac{\partial\phi}{\partial t} + \mathbb{F}_s|\nabla\phi|=0.
  \end{equation}

\noindent It is noted that Eq. (\ref{ls_advect_fs}) holds not only for $\phi=0$ but for all values of the level set function. Given the two equations for level set function, we are motivated to modify on the level set equation (\ref{ls_advect}) by selecting the term $\mathbb{F}_s|\nabla\phi|$ as the building block to reconstruct Eq. (\ref{ls_advect}) so as to avoid mass imbalance in the cells containing only the interface. Therefore, our proposed mass-preserving level set equation in cells with and without interface separating two different fluids is reconstructed as
   \begin{equation}
   	\label{mass-preserving ls}
   	\frac{\partial\phi}{\partial t}+\mathbf{U}\cdot\nabla\phi=\lambda_I\delta(\phi)|\nabla\phi|.
   \end{equation}
    It is worth to address here that Eq.(\ref{mass-preserving ls}) can be expressed differently by
    \begin{equation}
    	\label{mpls}
     		\frac{\partial\phi}{\partial t}+\mathbf{U}^c\cdot\nabla\phi=0,
    \end{equation}
    where $\mathbf{U}^c$ is the correction velocity
    \begin{equation}
    \label{correction velocity}
    	\mathbf{U}^c=\mathbf{U}-\lambda_I\delta(\phi)\frac{\nabla\phi}{|\nabla\phi|}.
    \end{equation}
  
It is now clear from Eq. (\ref{mpls}) and (\ref{correction velocity}) that our strategy to preserve the mass is to modify the velocity normally on the interface with a time-dependent parameter $\lambda_I$. As shown in Eq. (\ref{mass-preserving ls}),  $|\nabla\phi|$ introduced to the classical level set equation is the building block of rendering mass conservation property in the vicinity of the interface.  Inclusion of the term $|\nabla\phi|$ makes sense mathematically since $|\nabla\phi(\mathbf{{x}})|$ gives the shortest distance from $\mathbf{{x}}$ to the interface $\phi=0$. The coefficient $\lambda_I$ shown in the right-hand-side of (\ref{mass-preserving ls}) is a function of the evolving interface geometry, and it will be derived in detail in the next section. Delta function $\delta(\phi)$, which is the function of $\phi$ introduced in Eq. (\ref{mass-preserving ls}), is related to the Heaviside function $H(\phi)$ as follows
   \begin{equation}
   \label{delta_function}
   	\delta(\phi)=\frac{dH(\phi)}{d\phi}.
   \end{equation}
   
   \subsection{Derivation of a mass-preserving level set equation}
   \subsubsection{Mass and volume in level set method}
   Before starting the derivation of our proposed mass-preserving level set equation as show in Eq. (\ref{mass-preserving ls}), it is worth to address here why we choose to preserve the mass $\mathbb{M}(\phi)$ instead of volume $\mathbb{V}(\phi)$ of level set function $\phi$, which is different from previous study of other researchers. In realistic, the mass is conserved if and only if volume is conserved for incompressible flow. However, from the numerical point of view, the degree of sensitivity of volume $\mathbb{V}(\phi)$ and mass  $\mathbb{M}(\phi)$ are different to the level set function $\phi$, that is, $\rho\frac{\partial\mathbb{V}(\phi)}{\partial\phi}\neq\frac{\partial\mathbb{M}(\phi)}{\partial\phi}$. As a result, the error of level set function $\phi$ will lead to different amount of error in total mass and total volume. The total volume $\mathbb{V}\ofp$ and total mass $\mathbb{M}\ofp$ of level set function can be expressed as follow
   \begin{align}
    \mathbb{V}(\phi) &= \int\limits_{\Omega} \delta\mathbb{V}\ofp~d\Omega = \int\limits_{\Omega} H(\phi)~d\Omega, \label{Volume}\\
   	\mathbb{M}(\phi) &= \int\limits_{\Omega} \delta\mathbb{M}\ofp~d\Omega = \int\limits_{\Omega} \rho(\phi)H(\phi)~d\Omega. \label{Mass}
   \end{align}
   
\noindent To express the jump of density across the interfaces in the computational domain, Heaviside function $H\ofp$ is introduced. Then, the term $\rho\ofp$ shown in Eq.(\ref{Mass}) can be expressed as the sum of its small unit, that is

\begin{equation}
	\label{density}
	\rho\ofp = \rho_1 H\ofp + \bigl(1-H\ofp\bigr)\rho_2.
\end{equation}

\noindent In the above equation, fluid in the region with positive sign of the level set function has the density $\rho_1$, and the density of $\rho_2$ is the fluid in the region with negative sign of the level set function. As it is mentioned previously, the degree of sensitivity to the error of level set function can be estimated by computing the derivatives with respect to $\phi$. The derivative for total volume $\mathbb{V}\ofp$ can be expressed as follows
   \begin{equation}
   \label{volume change}
   	  \frac{\partial\mathbb{V}(\phi)}{\partial\phi} = \frac{\partial}{\partial\phi}\int\limits_{\Omega} H(\phi)~d\Omega =\int\limits_{\Omega}\frac{\partial H(\phi)}{\partial\phi}~d\Omega=\int\limits_{\Omega}\delta(\phi)~d\Omega.
   \end{equation}
   
\noindent The derivative for total mass $\mathbb{M}\ofp$ is as follows by virtue of Eq. (\ref{Mass}) 
   
   \begin{equation}
   \label{mass change1}
   	\frac{\partial\mathbb{M}\ofp}{\partial\phi} = \int\limits_{\Omega} \biggl[H\ofp\frac{\partial\rho\ofp}{\partial\phi} + \rho\ofp\frac{\partial H\ofp}{\partial\phi}\biggr]~d\Omega.
   \end{equation}
   
\noindent The derivative for the total mass can be further expressed as follows by substituting Eq. (\ref{density}) into Eq. (\ref{mass change1}).
  
  \begin{equation}
  \label{mass change2}
  \begin{aligned}
  &\int\limits_{\Omega} \biggl[H\ofp\frac{\partial\rho\ofp}{\partial\phi} + \rho\ofp\frac{\partial H\ofp}{\partial\phi}\biggr]~d\Omega\\
  =&\int\limits_{\Omega} \biggl[H\ofp\frac{\partial}{\partial\phi}\bigl[\rho_1 H\ofp + \bigl(1-H\ofp\bigr)\rho_2\bigr]+\bigl[\rho_1 H\ofp + \bigl(1-H\ofp\bigr)\rho_2\bigr]\delta\ofp\biggr]~d\Omega\\
  =&\int\limits_{\Omega}\biggl[H\ofp\bigl[ \rho_1\delta\ofp-\rho_2\delta\ofp \bigr]+\bigl[\rho_1 H\ofp + \bigl(1-H\ofp\bigr)\rho_2\bigr]\delta\ofp  \biggr]~d\Omega\\
  =&\int\limits_{\Omega}\bigl[2H\ofp\delta\ofp(\rho_1-\rho_2)+\rho_2\delta\ofp\bigr]~d\Omega.
  \end{aligned}
  \end{equation}
  
Finally, the expression of derivative of the total mass can be futher simplified and expressed as follows using Eq. (\ref{volume change}) 

\begin{equation}
\label{MV1}
\frac{\partial\mathbb{M}\ofp}{\partial\phi} = \rho_2\frac{\partial\mathbb{V}\ofp}{\partial\phi} + 2\Delta\rho\int\limits_{\Omega} H\ofp\delta\ofp~d\Omega,
\end{equation}
  
\noindent where $\Delta\rho=\rho_1-\rho_2$ is the density difference of fluid 1 and 2.\\

Derivatives of the differential volume and mass can be derived by substituting $\partial\mathbb{V}/\partial\phi = \int\limits_{\Omega} \partial(\delta\mathbb{V})/\partial\phi~d\Omega$ and $\partial\mathbb{M}/\partial\phi = \int\limits_{\Omega} \partial(\delta\mathbb{M})/\partial\phi~d\Omega$ into Eq. (\ref{MV1})

\begin{equation}
\label{MV2}
\frac{\partial(\delta\mathbb{M})}{\partial\phi} = \rho_2\frac{\partial(\delta\mathbb{V})}{\partial\phi} + 2\Delta\rho H\ofp\delta\ofp
\end{equation}

\noindent Let's assume there is a non-uniform distribution of an error, or $\Delta_\phi$, for the level set function. By multiplying $\Delta_\phi$ on the both side of Eq.(\ref{MV2}) and integrating the equation on the computational domain, we can then obtain

\begin{equation}
\int\limits_{\Omega}\frac{\partial(\delta\mathbb{M})}{\partial\phi}\cdot\Delta_\phi~d\Omega = \rho_2\int\limits_{\Omega} \frac{\partial(\delta\mathbb{V})}{\partial\phi}\cdot\Delta_\phi~d\Omega + 2\Delta\rho \int\limits_{\Omega} H\ofp\delta\ofp\cdot\Delta_\phi~d\Omega
\end{equation}

By the definition of the derivatives, the relation of error for the total mass and the error of the total volume can be expressed as

\begin{equation}
\label{MassVol diff}
\Delta\mathbb{M} = \rho_2\Delta\mathbb{V} + 2\Delta\rho \int\limits_{\Omega} H\ofp\delta\ofp\cdot\Delta_\phi~d\Omega
\end{equation}

In the light of the above equation, we are led to know that if the volume and the mass can be preserved simultaneously if the density is uniform in the computational domain. However, when there is density difference across the interface, the introduction of Heaviside function will cause an additional error term to appear. As a result, total mass and total volume are not necessarily to be conserved at the same time in two-phase flow simulations. Owing to the presence of the last term in Eq. (\ref{MassVol diff}), it is clear that preservation only to volume does not mean the preservation of mass when predicting two-phase flow. Due to the above reasons, it is more intuitive to preserve the mass instead of volume in practical simulations.\\
   
   \subsubsection{Mass-preserving level set equation}
   The total mass of a control volume $\Omega(t)$ at $t=n\Delta t$ can be expressed as follows,
   
   \begin{equation}
   	\mathbb{M}(\phi) = \int\limits_{\Omega(t)}\rho\ofp H(\phi)~d\Omega.
   \end{equation}
   
\noindent In Eulerian description, the expression of the rate change of $\mathbb{M}(\phi)$ can be derived by performing the total derivative on it 
   \begin{equation}
      \label{rtt_mass}
      \frac{d}{dt}\int\limits_{\Omega(t)}\rho\ofp H(\phi)~d\Omega = \int\limits_{\partial\Omega(t)}\rho\ofp H(\phi)\mathbf{U}\cdot\hat{n}~d(\partial\Omega) + \int\limits_{\Omega(t)}\frac{\partial}{\partial t}\biggl(\rho\ofp H(\phi)\biggr)~d\Omega.
   \end{equation}
   
\noindent In the above equation, $\partial\Omega(t)$ is the surface that encloses $\Omega(t)$. By employing Gauss' theorem, we are led to have $\int\limits_{\partial\Omega(t)}\rho\ofp H(\phi)\mathbf{U}\cdot\hat{n}~d\partial\Omega = \int\limits_{\Omega(t)}\nabla\cdot\bigl(\rho\ofp H(\phi)\mathbf{U}\bigr)~d\Omega$. Then, Eq. (\ref{rtt_mass}) can be further simplified as 
   
  \begin{equation}
    \label{rtt_mass2}
  	\frac{d}{dt}\int\limits_{\Omega(t)}\rho\ofp H(\phi)~d\Omega = \int\limits_{\Omega(t)}\biggl[\frac{\partial}{\partial t}\biggl(\rho\ofp H(\phi)\biggr)+\nabla\cdot\biggl(\rho\ofp H(\phi)\mathbf{U}\biggr)\biggr]~d\Omega.
  \end{equation}
   
\noindent By using the chain rule, one can expand Eq. (\ref{rtt_mass2}) to yield
  
  \begin{equation}
   \label{rtt_mass3}
   \begin{aligned}
  	&\frac{d}{dt}\int\limits_{\Omega(t)}\rho\ofp H(\phi)~d\Omega \\
  	&=\int\limits_{\Omega(t)}\biggl[\frac{\partial}{\partial t}\biggl(\rho\ofp H(\phi)\biggr)+\nabla\cdot\biggl(\rho\ofp H(\phi)\mathbf{U}\biggr)\biggr]~d\Omega\\  
  	&=\int\limits_{\Omega(t)}\biggl[H(\phi)\frac{\partial\rho\ofp}{\partial t}+\rho\ofp\frac{\partial H(\phi)}{\partial t}\biggr] + \biggl[\rho\ofp H(\phi)\nabla\cdot\mathbf{U}+\mathbf{U}\cdot\nabla\biggl(\rho\ofp H(\phi)\biggr)\biggr]~d\Omega.
  	\end{aligned}
  \end{equation}

\noindent For the case of incompressible flow, the term $\rho\ofp H(\phi)\nabla\cdot\mathbf{U}$ can be neglected, then Eq. (\ref{rtt_mass3}) can be further rewritten as follows with the substitution of the two identity, $\partial_tH\ofp=\delta(\phi)\partial_t\phi$ and $\mathbf{U}\cdot\nabla\bigl(\rho\ofp H\ofp\bigr)=\rho\ofp\delta(\phi)\mathbf{U}\cdot\nabla\phi+H\ofp\mathbf{U}\cdot\nabla\rho\ofp$. Recollection of the terms $H\ofp$ and $\rho\ofp\delta(\phi)$, the following equation is yielded

\begin{equation}
\begin{aligned}
&\frac{d}{dt}\int\limits_{\Omega(t)}\rho\ofp H(\phi)~d\Omega\\
&= \int\limits_{\Omega(t)}\biggl[H(\phi)\frac{\partial\rho\ofp}{\partial t}+\rho\ofp\frac{\partial H(\phi)}{\partial t}\biggr] + \rho\ofp\delta(\phi)\mathbf{U}\cdot\nabla\phi+H\ofp\mathbf{U}\cdot\nabla\rho\ofp~d\Omega\\
&=\int\limits_{\Omega(t)} H\ofp\biggl[\frac{\partial\rho\ofp}{\partial t}+\mathbf{U}\cdot\nabla\rho\ofp\biggr] + \rho\ofp\delta(\phi)\biggl[\frac{\partial\phi}{\partial t}+\mathbf{U}\cdot\nabla\phi\biggr]~d\Omega\\
&=\int\limits_{\Omega(t)} H\ofp\frac{d\rho\ofp}{dt} + \rho\ofp\delta\ofp\frac{d\phi}{dt}~d\Omega
\end{aligned}
\end{equation}

\noindent Next, using the identity $d\rho\ofp/dt=(\partial\rho/\partial\phi) \cdot (d\phi/dt)$ and the equation of density shown in Eq. (\ref{density}), we can then obtain
 
 \begin{equation}\label{mass change time}
 \begin{aligned}
 &\frac{d}{dt}\int\limits_{\Omega(t)}\rho\ofp H(\phi)~d\Omega\\
 &=\int\limits_{\Omega(t)} \biggl[ H\ofp\frac{\partial\rho\ofp}{\partial\phi}+ \rho\ofp\delta\ofp\biggr]\frac{d\phi}{dt}~d\Omega\\
 &=\int\limits_{\Omega(t)} \frac{\partial}{\partial\phi}\bigl[H\ofp\rho\ofp\bigr]\cdot\frac{d\phi}{dt}~d\Omega
 \end{aligned}
 \end{equation}
 
\noindent As it is shown in above equation, different selection of numerical smooth Heaviside function $H\ofp$ will have different effect on the rate change of total mass. Using Eq. (\ref{density}), the rate change of mass in a control volume $\Omega(t)$ can be expressed as follows
 
 \begin{equation}
 \label{mass change time 2}
 	\frac{d}{dt}\int\limits_{\Omega(t)}\rho\ofp H(\phi)~d\Omega = \int\limits_{\Omega(t)} \delta\ofp\biggl[2(\rho_1-\rho_2) H\ofp+{\rho_2}\biggr]\frac{d\phi}{dt}~d\Omega
 \end{equation}
   
    It is apparent from Eq. (\ref{mass change time 2}) that the non-conservation of mass is attributed to two different types of errors on the interface. The first part is mainly caused by the introduction smooth Heaviside function $H\ofp$, and this error is weighted by the density difference $\Delta\rho=\rho_1-\rho_2$. The second error leading to loss of mass origins from the indispensable discretization error $\mu_\phi$ introduced in the approximation of Eq. (\ref{ls_advect}). It is worth to address here that the presence of $\mu_\phi$ is mainly caused by the discretization error introduced from the approximation of the term $\mathbf{U}\cdot\nabla\phi$. Application of different numerical schemes will generate different values of $\mu_\phi$. Our proposed mass-preserving level set model is rooted in the introduction of a proper source into Eq. (\ref{ls_advect}) to dispense the error $\mu_\phi$ and to preserve the total mass of the control volume. \\
     
     According to the above derivation, the rate change of mass of classical level set equation (\ref{ls_advect}) can then be written as
     \begin{equation}
        \label{mass-change-ls}
     	\frac{d}{dt}\bigl(\mathbb{M}^{LS}(\phi)\bigr) = \int\limits_{\Omega(t)} \delta\ofp\biggl[2(\rho_1-\rho_2) H\ofp+{\rho_2}\biggr]\mu_\phi~d\Omega
     \end{equation}
     Similarly, the rate of change of the total mass in our proposed mass-preserving level set method can be derived as follows by substituting Eq. (\ref{mass-preserving ls}) into Eq. (\ref{mass change time 2})
    \begin{equation}
     \label{rtt_mass_new}
     \begin{aligned}
     \frac{d}{dt}\bigl(\mathbb{M}^{MPLS}(\phi)\bigr) &= \int\limits_{\Omega(t)} \delta\ofp\biggl[2(\rho_1-\rho_2) H\ofp+{\rho_2}\biggr]\biggl[\mu_\phi+\lambda_I\delta\ofp|\nabla\phi|\biggr]~d\Omega\\
     &=\frac{d}{dt}\bigl(\mathbb{M}^{LS}(\phi)\bigr) + \lambda_I\int\limits_{\Omega(t)}\delta^2\ofp|\nabla\phi|\biggl[2(\rho_1-\rho_2) H\ofp+{\rho_2}\biggr]~d\Omega
     \end{aligned}
    \end{equation}
    Our goal is to preserve the total mass of the level set function in the course of simulation. As a result, the coefficient $\lambda_I$ can then be derived by imposing the condition $d\bigl(\mathbb{M}^{MPLS}(\phi)\bigr)/dt=0$, thereby leading to 
    \begin{equation}
        \label{lamb 1}
    	\lambda_I = -\frac{d\bigl(\mathbb{M}^{LS}(\phi^n)\bigr)/dt}{\int\limits_{\Omega(t)}\delta^2\ofp|\nabla\phi|\bigl[2(\rho_1-\rho_2) H\ofp+{\rho_2}\bigr]~d\Omega} .
    \end{equation}

    \subsubsection{Numerical Implementation}
   Eq. (\ref{lamb 1}) is the expression that makes the rate change of the total mass of  the proposed mass-preserving level set equation to be zero. However, from the numerical point of view, a direct calculations of Eq. (\ref{mass-change-ls}) may deteriorate the computational efficiency significantly in the simulation. We are therefore motivated to measure the rate change of mass as the growth of difference between numerical solution and the exact solution. The exact value of total mass can be expressed below
   
   \begin{equation}
   \label{exact mass}
   \mathbb{M}_{exact}(t) = \mathbb{M}_0 + \int^t_0\iint\limits_{\partial\Omega(t)}\mathbf{\Psi}(t)\cdot\hat{n}~d(\partial\Omega)\;dt
   \end{equation}
   where $\mathbf{\Psi}$ is the mass flux across the surface $\partial\Omega(t)$ that encloses the control volume $\Omega(t)$, and $\mathbb{M}_0$ is the total mass at $t=0$. Therefore, the parameter $\lambda_I$ can be expressed as follows
   
   \begin{equation}
   \label{lambda 1}
   	\lambda_I = \frac{\mathbb{M}_{exact} - \mathbb{M}^{LS}(\phi^{n+1})}{\Delta t\int\limits_{\Omega(t)}\delta^2\ofp|\nabla\phi|\bigl[2(\rho_1-\rho_2) H\ofp+{\rho_2}\bigr]~d\Omega}
   \end{equation}
   To implement our proposed method directly, it is recommended to split the the mass-preserving level set equation Eq. (\ref{mass-preserving ls}) into two parts. The first part is identical to the classical level set equation Eq. (\ref{ls_advect}), and the second part is mass-preserving correction step, given below in Eq. (\ref{correction step}). After the Eq. (\ref{ls_advect}), the value of $\mathbb{M}^{LS}(\phi^{n+1})$ can then be calculated. Thereby, the mass-preserving correction step can be directly carried out.
   
   \begin{equation}
   \label{correction step}
      \frac{\partial\phi}{\partial t} = \lambda_I\delta\ofp|\nabla\phi|
    \end{equation}    
    
It is noted that the density $\rho$ is usually normalized to the form of dimensionless, given in Eq. (\ref{rhomu_norm}). Under the circumstance, the parameter $\lambda_I$ shown in Eq. (\ref{lambda 1}) is replaced with following equation

\begin{equation}
\label{lambda 2}
\lambda_I = \frac{\mathbb{M}_{exact} - \mathbb{M}^{LS}(\phi^{n+1})}{\Delta t\int\limits_{\Omega(t)}\delta^2\ofp|\nabla\phi|\bigl[2(1-\rho_{12}) H\ofp+{\rho_{12}}\bigr]~d\Omega}.
\end{equation}
In the above equation, $\rho_{12}=\rho_2/\rho_1$ is the density ratio of fluid 1 and 2.
   \subsection{Re-initialization of mass-preserving level set equation}
   
   The level set advection equation does not necessarily require that the level set function $\phi$ be a distance function. In an implicit representation of an interface, we demand, however, that the chosen level set function $\phi$ in Eq. (\ref{mass-preserving ls}) be a signed distance function with the property of $|\nabla\phi|=1$ from the numerical point of view. If $\phi$ is not a distance function, numerical approximation can be quite inaccurate provided that $\phi$ has a large variation in its gradient. In this study, we therefore keep $\phi$ close to the signed distance function through a frequent application of the reinitialization procedure.\\
   
    Reinitialization can be achieved by solving the following Hamilton-Jacobi equation proposed in \cite{Sussman(1999),Rouy(1992)} to reconstruct the level set function with the exact zero isovalue of $\phi(\mathbf{x})$.
   
   \begin{equation}
   \label{reini}
   \begin{aligned}
   	&\frac{\partial\phi^*}{\partial\tau} + \bar{S}(\phi^*_0)\bigl(|\nabla\phi^*|-1\bigr)=\lambda_R \delta(\phi^*)|\nabla\phi^*|,\\
   	&\phi^*(\tiny{\tau=0},\mathbf{x}) = \phi^*_0 = \phi(t,\mathbf{x}),
   	\end{aligned}
   \end{equation}
   
   where the time-dependent parameter $\lambda_R$ is given as \cite{Sussman(1999)}
   
   \begin{equation}
   	\lambda_R=\frac{-\int_{\Omega_{i,j,k}}\delta(\phi^*)\bar{S}(\phi^*_0)\bigl(1-|\nabla\phi^*|\bigr)d\Omega}{\int_{\Omega_{i,j,k}}\delta^2(\phi^*)|\nabla\phi^*| d\Omega}.
   \end{equation}
   In the above equations, $\phi^*_0$ is the level set function prior to performing re-initialization and the virtual time $\tau$ is introduced for iteration purpose. In Eq. (\ref{reini}), $\bar{S}(\phi^*_0)$ is the signed distance function defined as
   
   \begin{equation}
\label{sign_function_exact}
\bar{S}(\phi^*_0)=\left\{
\begin{array}{ll}   
1;  &\mbox{if }\phi^*_0>0,\\
0;  &\mbox{if }\phi^*_0=0,\\
-1;  &\mbox{if }\phi^*_0<0.\\
\end{array}
\right.
   \end{equation}
    One can also choose $\bar{S}(\phi^*_0)$ as 
   
   \begin{equation}
      \label{sign_function}
   	  \bar{S}(\phi^*_0) = 2H(\phi^*_0)-1.
   \end{equation} 
   
   It is worthy to note that Eq. (\ref{reini}) can be expressed differently as follows
   \begin{equation}
   	\frac{\partial\phi^*}{\partial\tau} + \mathbf{U}_r\cdot\nabla\phi^* = \bar{S}(\phi^*_0)+\lambda_R \delta(\phi^*)|\nabla\phi^*|,
   \end{equation}
   
\noindent where 
   
   \begin{equation}
   \mathbf{U}_r = \bar{S}(\phi^*_0)\rfrac{\nabla\phi^*}{|\nabla\phi^*|}.
   \end{equation}

\section{Mathematical model for incompressible two-phase flow simulation}
In this study we are aimed to simulate incompressible two-phase flow motion that incorporates surface tension force along the moving interface separating two different fluids. Our strategy of conducting the current simulation is to divide the whole flow domain into two sub-domains filled with individual phases or fluid media. Some physical properties such as density and viscosity are discontinuous across the interface between two sub-domains. One-field formulation will be adopted in this study by smoothing these physical properties over a transition region of fairly small finite thickness.\\

With the assumption that the fluid properties are constant in both sub-domains, the mass and momentum conservation equations for the incompressible Newtonian fluid flows can be written as

\begin{align}
  \label{continuity}
  \nabla\cdot\mathbf{U}=0,\\
  \label{momentum}
  \mathbf{U}_t+(\mathbf{U}\cdot\nabla)\mathbf{U}=-\frac{1}{\rho}\nabla p + g\hat{\mathbf{e}}_g + \frac{1}{\rho}\nabla\cdot\bigl(2\mu\underline{\mathbf{D}}\bigr) - \frac{\sigma\kappa}{\rho}\nabla H.
\end{align}

In the above momentum equation containing a surface tension force, $\hat{n}$ is the unit normal vector at the interface, $\kappa$ is the curvature of the interface, $H$ is the Heaviside function, $\underline{\mathbf{D}}~(\equiv\rfrac{1}{2}\bigl[\nabla\mathbf{U}+(\nabla\mathbf{U})^T\bigr])$ is the rate of deformation tensor, and $\sigma$ is the surface tension coefficient.\\

 In our one-field formulation of two-phase flow, the effective density $\rho$ and viscosity $\mu$ shown in equation (\ref{momentum}) at each grid point are approximated as follows for fluids 1 and 2:

\begin{align}
\label{rho}
\rho = \rho_2 \bigl(1-H\bigr) + \rho_1 H, \\
\label{mu}
\mu = \mu_2 \bigl(1-H\bigr) + \mu_1 H,
\end{align}

\noindent where $H$ denotes the smooth Heaviside function. Note that $H$ is introduced for the purpose of preventing numerical instability arising from the steep gradients of $\rho$ and $\mu$. The subscripts 1 and 2 represent fluid 1 and 2, respectively. The smoothed Heaviside function \cite{Sussman(1994)} employed in this study is defined as the function of level set function $\phi$:

\begin{equation}
\label{heaviside}
{H}(\phi)=\left\{
\begin{aligned}   
&0&;\mbox{if }\phi<-\varepsilon,\\
&\frac{1}{2}\biggl[1+\frac{\phi}{\varepsilon}+\frac{1}{\pi}\sin(\frac{\pi\phi}{\varepsilon})\biggr]&;\mbox{if }|\phi|\leq \varepsilon,\\
&1&;\mbox{if }\phi>\varepsilon.\\
\end{aligned}
\right.
\end{equation}
In this study, the numerical interface thickness $\varepsilon$ is chosen to be equal to one and half of the size of a cell (or $\varepsilon=1.5\Delta x$). It is noted that the corresponding smoothed delta function in Eq. (\ref{delta_function}) and sign function in Eq. (\ref{sign_function}) can be obtained directly through their definitions associated with $H(\phi)$ in Eq. (\ref{heaviside}). Moreover, $\nabla H$ shown in Eq. (\ref{momentum}) can be replaced by $\delta(\phi)\nabla\phi$. Although the level set method is less attractive to get the same level of the conservation property as the VOF method or front tracking method, we adopt this interface capturing method due to an inevitable presence of the  geometric quantities $\hat{n}$ and $\kappa$ in Eq. (\ref{momentum}). The main reason lies in its inherent good ability to compute $\hat{n}$ and $\kappa$ more accurately through the predicted level set values of $\phi$ by means of

 \begin{align}
 \hat{n} &= \frac{\nabla\phi}{|\nabla\phi|} = \frac{\phi_x\hat i+\phi_y\hat j}{\sqrt{\phi_x^2+\phi_y^2}},\\
 \kappa &= \nabla\cdot\hat{n} = \frac{\phi_x^2\phi_{yy}-2\phi_x\phi_y\phi_{xy}+\phi_y^2\phi_{xx}}{(\phi_x^2+\phi_y^2)^{3/2}}.
 \end{align}
 In the three-dimensional case, $\hat{n}$ and $\kappa$ can be expressed as

\begin{align}
\hat{n} &= \frac{\nabla\phi}{|\nabla\phi|} = \frac{\phi_x\hat i+\phi_y\hat j+\phi_z \hat k}{\sqrt{\phi_x^2+\phi_y^2+\phi_z^2}},\\
\kappa &= \nabla\cdot\hat{n} = \frac{\phi_x^2(\phi_{yy}+\phi_{zz})+\phi_y^2(\phi_{xx}+\phi_{zz})+\phi_z^2(\phi_{xx}+\phi_{yy})-2(\phi_x\phi_y\phi_{xy}+\phi_x\phi_z\phi_{xz}+\phi_y\phi_z\phi_{yz})}{(\phi_x^2+\phi_y^2)^{3/2}}.
\end{align}
Note that application of the VOF method will be enormously difficulty to compute an accurate value of the local curvature from the predicted volume fraction near the interface.

\subsection{Normalization of equations}

Dimensional analysis is a widely-used technique in fluid mechanics with the following characteristic values: $L_c$, the characteristic length; $U_c$, the characteristic velocity; $\Gamma=\rho_1 U_c^2$, the characteristic pressure; $T_c=L_c/U_c$, the characteristic time; $\rho_1$, the characteristic density; and $\mu_1$, the characteristic viscosity. By virtue of the above scalings to normalize Eqs. (\ref{continuity}) and (\ref{momentum}), one can derive the following dimensionless equations from Eq. (\ref{continuity}) and Eq. (\ref{momentum}), respectively:

\begin{align}
    \label{con-norm}
	\nabla\cdot\mathbf{U}=0,\\
	\label{NS-norm}
	\mathbf{U}_t + (\mathbf{U}\cdot\nabla)\mathbf{U} = -\frac{1}{\rho(\phi)}\nabla p + \frac{1}{Re}\frac{\nabla\cdot\bigl(2\mu(\phi)\underline{\mathbf{D}}\bigr)}{\rho(\phi)}+\frac{1}{Fr^2}\hat{\mathbf{e}_g} - \frac{1}{We}\frac{\kappa(\phi)\delta(\phi)\nabla\phi}{\rho(\phi)}.
\end{align} 

The dimensionless density and viscosity are as follows

\begin{equation}
\label{rhomu_norm}
\begin{aligned}
	&\rho(\phi) = H(\phi) + \bigl(1-H(\phi)\bigr)\frac{\rho_2}{\rho_1},\\
	&\mu(\phi) = H(\phi) + \bigl(1-H(\phi)\bigr)\frac{\mu_2}{\mu_1}.
\end{aligned}
\end{equation}

The direct consequence of the  above normalization of equations is the introduction of three dimensionless parameters, which are the Reynolds number $Re\equiv{\rho_1L_cU_c}/{\mu_{L}}$ for representing the ratio of the inertial force to viscous force of the fluid, 
the Weber number $We\equiv{\rho_{1}L_cU_c^{2}}/{\sigma}$ for representing the ratio of the inertial force to gravity force of the fluid, 
and the Froude number $Fr\equiv{U_c}/{\sqrt{gL_c}}$ for representing the ratio of the inertial force to gravitational force of the fluid. 

\section{Numerical models}

\subsection{Spatial approximation of advection equation}

To get a smaller absolute truncation error, a scheme with higher-order accuracy and with smaller dispersion error in smooth regions shall be chosen. We also aim to avoid discontinuous solutions near discontinuities. For achieving the above two goals simultaneously, the optimized compact reconstruction weighted essentially non-oscillatory (OCRWENO4) scheme \cite{Gu(2018)} is applied for the convective flux term shown in the proposed mass-preserving level set equation.\\

 In two-dimensional space, approximation of the convective flux term in Eq. (\ref{mass-preserving ls}) can be written in its conservative form as follows for an incompressible fluid flow

\begin{equation}
\label{ls_advect_flux_discrete}
\mathbf{U}\cdot\nabla\phi = \nabla\cdot\bigl(\mathbf{U}\phi\bigr) = \frac{F_{i+1/2,j}-F_{i-1/2,j}}{\Delta x} + \frac{G_{i+1/2,j}-G_{i-1/2,j}}{\Delta y}.
\end{equation}

In the above equation, $F_{i+1/2,j}$ and $G_{i,j+1/2}$ are the numerical fluxes reconstructed at the cell face along $x,y$ direction, respectively.\\

Reconstruction of convective fluxes lies in the use of Lax-Friedrichs splitting method \cite{Crandall(1984)} such that the term $F_{i+1/2,j}$ can be written as follows

\begin{equation}
F_{i+1/2,j} = \frac{1}{2}\biggl(\breve F^L_{i+1/2,j}+\hat F^R_{i+1/2,j}\biggr) =  \frac{1}{2}\biggl((u^+\phi)^L_{i+1/2,j}+(u^-\phi)^R_{i+1/2,j}\biggr).
\end{equation}

The expression of $G_{i,j+1/2}$ can be derived similarly as well. Note that $u^+=u+|u|$ and $u^-=u-|u|$, and the subscripts $L,R$ denote the reconstruction of OCRWENO4 scheme from the left and right biased interpolations, respectively. The value of $\breve{F}^L_{i+1/2,j}=(u^+\phi)^L_{i+1/2,j}$ can be obtained by solving the following  tridiagonal matrix equation \cite{Ghosh(2012)}

\begin{equation}	
\label{ocrweno-left}
\begin{aligned}
\biggl[\frac{2\omega^L_1+\omega^L_2}{3}\biggr]\breve{F}^L_{i-\frac{1}{2}} + \biggl[\frac{\omega^L_1+2(\omega^L_2+\omega^L_3)}{3}\biggr]\breve{F}^L_{i+\frac{1}{2}}+\frac{\omega^L_3}{3}\breve{F}^L_{i+\frac{3}{2}}\\ 
=\frac{\omega^L_1}{6}\breve{F}_{i-1}+\biggl[\frac{5(\omega^L_1+\omega^L_2)+\omega^L_3}{6}\biggr]\breve{F}_i+\biggl[\frac{\omega^L_2+5\omega^L_3}{6}\biggr]\breve{F}_{i+1}.
\end{aligned}
\end{equation}

In the above equation, $\omega^L_k, k=1,2,3$, are the weighting factors associated with the smoothness indicators $\beta^L_k,k=1,2,3$, to detect the degree of discontinuity in grid stencil to properly interpolate the numerical flux at cell face. Expressions of $\omega^L_k$ and $\beta^L_k$ are given as follows

\begin{equation}
\label{factor_left}
\begin{aligned}
&\omega^L_k = \frac{\alpha^L_k}{\Sigma_k~\alpha^L_k},~\alpha^L_k = c_k \biggl(1+\frac{|\beta^L_3 - \beta^L_1|}{\epsilon + \beta^L_i}\biggr),\\
\beta^L_1&=\frac{13}{12}{(\breve{F}_{i-2}-2\breve{F}_{i-1}+\breve{F}_i)}^2+\frac{1}{4}{(\breve{F}_{i-2}-4\breve{F}_{i-1}+3\breve{F}_i)}^2,\\
\beta^L_2&=\frac{13}{12}{(\breve{F}_{i-1}-2\breve{F}_{i}+\breve{F}_{i+1})}^2+\frac{1}{4}{(\breve{F}_{i-1}-\breve{F}_{i+1})}^2,\\
\beta^L_3&=\frac{13}{12}{(\breve{F}_{i}-2\breve{F}_{i+1}+\breve{F}_{i+2})}^2+\frac{1}{4}{(3\breve{F}_{i}-4\breve{F}_{i+1}+\breve{F}_{i+2})}^2. 
\end{aligned}
\end{equation}

The corresponding tridiagonal matrix equation for $\hat{F}^R_{i+1/2,j}=(u^-\phi)^R_{i+1/2,j}$ is given below

\begin{equation}	
\label{ocrweno-right}
\begin{aligned}
\biggl[\frac{2\omega^R_1+\omega^R_2}{3}\biggr]\hat{F}^R_{i+\frac{3}{2}} + \biggl[\frac{\omega^R_1+2(\omega^R_2+\omega^R_3)}{3}\biggr]\hat{F}^R_{i+\frac{1}{2}}+\frac{\omega^R_3}{3}\hat{F}^R_{i-\frac{1}{2}}\\ 
=\frac{\omega^R_1}{6}\hat F_{i+2}+\biggl[\frac{5(\omega^R_1+\omega^R_2)+\omega^R_3}{6}\biggr]\hat F_{i+1}+\biggl[\frac{\omega^R_2+5\omega^R_3}{6}\biggr]\hat F_{i}.
\end{aligned}
\end{equation}

The weighting factors associated with the smoothness indicators are given as

\begin{equation}
\label{factor_right}
\begin{aligned}
&\omega^R_k = \frac{\alpha^R_k}{\Sigma_k~\alpha^R_k},~\alpha^R_k = c_k \biggl(1+\frac{|\beta^R_3 - \beta^R_1|}{\epsilon + \beta^R_k}\biggr),\\
\beta^R_1&=\frac{13}{12}{(\hat F_{i+1}-2\hat F_{i+2}+\hat F_{i+3})}^2+\frac{1}{4}{(3\hat F_{i+1}-4\hat F_{i+2}+\hat F_{i+3})}^2,\\
\beta^R_2&=\frac{13}{12}{(\hat F_{i}-2\hat F_{i+1}+\hat F_{i+2})}^2+\frac{1}{4}{(\hat F_{i}-\hat F_{i+2})}^2,\\
\beta^R_3&=\frac{13}{12}{(\hat F_{i-1}-2\hat F_{i}+\hat F_{i+1})}^2+\frac{1}{4}{(\hat F_{i-1}-4\hat F_{i}+3\hat F_{i+1})}^2.
\end{aligned}
\end{equation}

The magnitude of the parameter $\epsilon$ shown in (\ref{factor_left}) and (\ref{factor_right}) is set at $10^{-8}$ to avoid zero-valued denominator. The optimized coefficients shown in (\ref{factor_left}) and (\ref{factor_right}) are given by  $c_{1}=0.20891413$, $c_{2}=0.49999999$ and $c_{3}=0.29108586$ which altogether can yield a fourth order accuracy approximation with low dispersion error in the approximation for spatial derivatives. One can refer to \cite{Ghosh(2012),Gu(2018)} for the detailed derivation of the optimized coefficients.

\subsection{Spatial approximation of re-initialization equation}

Eq. (\ref{reini}) can be rewritten in the following Hamilton-Jacobi form:

\begin{equation}
\label{reini_hamil}
\phi_\tau + \bar{S}(\phi_0)\mathbb{H}(\phi,\nabla\phi) =0.
\end{equation}

Note that the subscript $*$ has been omitted for convenience. In the above equation, $\mathbb{H}(\phi,\nabla\phi)=|\nabla\phi|-1$ is the corresponding Hamiltonian function. By employing Godunov spatial discretization given in \cite{Osher(1989icase)}, Eq. (\ref{reini_hamil}) can be further rewritten in terms of the one-sided derivatives $\phi_x^L,\phi_x^R,\phi_y^L,\phi_y^R$,

\begin{equation}
\label{reini_hamil_god}
\phi_\tau + \bar{S}(\phi_0)\mathbb{H}^G(\phi_x^L,\phi_x^R,\phi_y^L,\phi_y^R) =0
\end{equation}

In the above equation, $\mathbb{H}^G$ is a function of the one-sided derivatives defined as follows

\begin{equation}
\label{god}
\mathbb{H}^G(\phi_x^L,\phi_x^R,\phi_y^L,\phi_y^R) =
\left\{\begin{array}{cc}
&\mbox{if $\bar{S}(\phi_0)\leq0$},\\
&\sqrt{\max{\bigl[((\phi_x^R)^p)^2,((\phi_x^L)^m)^2\bigr]}+\max{\bigl[((\phi_y^R)^p)^2,((\phi_y^L)^m)^2\bigr]}}-1,\\
&\mbox{if $\bar{S}(\phi_0)>0$},\\
&\sqrt{\max{\bigl[((\phi_x^R)^m)^2,((\phi_x^L)^p)^2\bigr]}+\max{\bigl[((\phi_y^R)^m)^2,((\phi_y^L)^p)^2\bigr]}}-1.
\end{array} \right.
\end{equation}

It si noted that the notations $(\cdot)^p=\max(\cdot~,0)$ and $(\cdot)^m=\min(\cdot~,0)$ have been applied to the above equation.\\

 Calculations of the one-sided derivatives $\phi_x^L,\phi_x^R,\phi_y^L,\phi_y^R$ follow the classical WENO5 scheme presented in \cite{Shu(1997icase)}. Take for example a one-dimensional case, $\phi_x^L,\phi_x^R$ can be expressed as follows

\begin{equation}
\begin{aligned}
\phi_x^L\bigr|_i  &= \frac{\phi^L_{i+1/2}-\phi^L_{i-1/2}}{x_{i+1/2}-x_{i-1/2}},\\
\phi_x^R\bigr|_i  &= \frac{\phi^R_{i+1/2}-\phi^R_{i-1/2}}{x_{i+1/2}-x_{i-1/2}}.
\end{aligned}
\end{equation}

Within the framework of WENO5 scheme, $\phi^L_{i+1/2}$ and $\phi^R_{i+1/2}$  can be approximated explicitly by using the following equations

\begin{equation}
\label{weno}
\begin{split}
\phi^{L}_{i+\frac{1}{2}}=&\;\frac{\omega_{1}}{3}\phi_{i-2}-\frac{1}{6}\biggl(7\omega_{1}+\omega_{2}\biggr)\phi_{i-1}+\frac{1}{6}\biggl(11\omega_{1}+5\omega_{2}+2\omega_{3}\biggr)\phi_{i}\\
&+\frac{1}{6}\biggl(2\omega_{2}+5\omega_{3}\biggr)\phi_{i+1}-\frac{\omega_{3}}{6}\phi_{i+2},\\
\phi^{R}_{i+\frac{1}{2}}=& -\frac{\widetilde{\omega}_3}{6}\phi_{i-1}+\frac{1}{6}\biggl(2\widetilde{\omega}_2+5\widetilde{\omega}_3\biggr)\phi_i+\frac{1}{6}\biggl(11\widetilde{\omega}_1+5\widetilde{\omega}_2+2\widetilde{\omega}_3\biggr)\phi_{i+1}\\
&-\frac{1}{6}\biggl(7\widetilde{\omega}_1+\widetilde{\omega}_2\biggr)\phi_{i+2}+\frac{\widetilde{\omega}_1}{3}\phi_{i+3},
\end{split}
\end{equation}

where the wighting factors $\omega_k$ and $\widetilde\omega_k$ with $k=1,2,3$ are given as:

\begin{equation}
\label{weno_weighting}
\begin{aligned}
\omega_k &= \frac{\alpha_k}{\Sigma_k~\alpha_k}, \alpha_k = \frac{\hat c_k}{\bigl(\beta^L_k+\epsilon\bigr)^2},\\
\widetilde\omega_k &= \frac{\widetilde\alpha_k}{\Sigma_k~\widetilde\alpha_k}, \widetilde\alpha_k = \frac{\hat c_k}{\bigl(\beta^R_k+\epsilon\bigr)^2}.
\end{aligned}
\end{equation}

The smoothness indicators $\beta^L_k,\beta^R_k$ can be obtained by replacing $\breve F, \hat F$ with $\phi$ in Eqs. (\ref{factor_left}) and (\ref{factor_right}). The optimal coefficients $\hat c_k$ shown in Eq. (\ref{weno_weighting}) are $\hat c_1 = 0.1,~\hat c_2 = 0.6,~\hat c_3=0.3$, which yield fifth order accuracy for the approximation of the one-sided derivatives.

\subsection{Temporal discretization method}
An explicit third-order Runge--Kutta (TVD-RK3) time discretization scheme \cite{Gottlieb(1998)} is used to solve Eqs. (\ref{mpls1}) and (\ref{reini}).
For example, both equations can be written as the following ODEs:
\begin{align}
\frac{d\phi}{d t} =L(\phi).
\end{align}
The TVD-RK3 scheme is then applied to yield the following three solution steps

\begin{equation}
\begin{aligned}
\phi^{(1)}&=\phi^{(n)}+\Delta t L(\phi^{(0)}),\\
\phi^{(2)}&=\frac{3}{4}\phi^{(n)}+\frac{1}{4}\phi^{(1)}+\frac{1}{4}\Delta t  L(\phi^{(1)}),\\
\phi^{(n+1)}&=\frac{1}{3}\phi^{(n)}+\frac{2}{3}\phi^{(2)}+\frac{2}{3}\Delta t  L(\phi^{(2)}).
\end{aligned}
\end{equation}

It is noted that we only use first order Euler scheme to solve the correction step Eq. (\ref{mpls2}).

\subsection{Navier-Stokes equation solver}
Based on the projection method \cite{Chorin(1968)}, the velocity can be obtained by using a four-step solution algorithm. Firstly, the pressure gradient term $\nabla p$ has been neglected to compute the intermediate velocity $\mathbf{U}^*$

\begin{equation}
\label{int_vel}
	\frac{\mathbf{U}^*-\mathbf{U}^n}{\Delta t} + \mathbf{S}^n = 0,
\end{equation}

where the source term $\mathbf{S}^n$ is approximated by the following explicit second-order Adams-Bashforth scheme

\begin{equation}
	\mathbf{S}^n = \frac{1}{2}\bigl(3\mathbf{A}^n-\mathbf{A}^{n-1}\bigr).
\end{equation}

In the above equation, the term $\mathbf{A}^n$ is the right-hand-side of Eq. (\ref{NS-norm}) without consideration of the pressure gradient, which can be expressed as

\begin{equation}
 \label{A}
  \mathbf{A}^n\equiv \bigl(\mathbf{U}^n\cdot\nabla\bigr)\mathbf{U}^n-\frac{1}{Re}\frac{\nabla\cdot\bigl(2\mu(\phi)\underline{\mathbf{D}}\bigr)}{\rho(\phi^n)}-\frac{1}{Fr^2}\hat{\mathbf{e}}_g+\frac{1}{We}\frac{\kappa(\phi^n)\delta(\phi^n)\nabla\phi^n}{\rho(\phi^n)}.
\end{equation}

In the calculation of $\mathbf{A}^n$, the term $\bigl(\mathbf{U}^n\cdot\nabla\bigr)\mathbf{U}^n$ is approximated by the third-order QUICK (quadratic upwind interpolation
for convective kinematics) scheme \cite{Leonard(1979)}. As for the diffusion term $\nabla\cdot\bigl(2\mu(\phi)\underline{\mathbf{D}}\bigr)$, it is approximated by the second-order central scheme, which can be expressed as follows, for example, for the one-dimensional case

\begin{equation}
\begin{aligned}
 &v\frac{\partial u}{\partial x}\biggr|_i =v_i\frac{u_{i+1/2}-u_{i-1/2}}{\Delta x},~u_{i+1/2}=\left\{
 \begin{array}{ll}   
&\frac{1}{8}(-u_{i-1}+6u_{i}+3u_{i+1}),~\mbox{if $v_{i+1/2}\geq 0$},\\
\\
&\frac{1}{8}(-u_{i+2}+6u_{i+1}+3u_{i}),~\mbox{if $v_{i+1/2}<0$},
 \end{array}
 \right.\\
 &\frac{\partial^2u}{\partial x^2}\biggr|_i = \frac{u_{i+1}-2u_i+u_{i-1}}{\Delta x^2}.
\end{aligned}
\end{equation}

After the intermediate velocity $\mathbf{U}^*$ is sought from Eq. (\ref{int_vel}), the velocity at $t=(n+1)\Delta t$ can be calculated by reconsidering the gradient of pressure to $\mathbf{U}^*$, which can be expressed as

\begin{equation} 
\label{vel_new}
	\frac{\mathbf{U}^{n+1}-\mathbf{U}^*}{\Delta t} = -\frac{1}{\rho^{n+1}}\nabla p^{n+1}
\end{equation}

In the above equation, the pressure value $p^{n+1}$ can be solved by performing the divergence operator on both sides of Eq. (\ref{vel_new}) with an imposed constraint $\nabla\cdot\mathbf{U}^{n+1}=0$. The pressure Poisson equation can then be derived as

\begin{equation}
    \label{poi}
	\nabla\cdot\biggl(\frac{1}{\rho^{n+1}}\nabla p^{n+1}\biggr) = \frac{\nabla\cdot\mathbf{U}^*}{\Delta t}.
\end{equation}

The pressure Poisson equation is solved by using the second-order central difference scheme and the point successive over-relaxation method. In two-dimensional space, the discretized expression of Eq. (\ref{poi}) is

\begin{equation}
\label{poi2}
Ap_{i-1,j}+Bp_{i+1,j}+Cp_{i,j}+Dp_{i,j+1}+Ep_{i,j-1}=\biggl(\frac{\nabla\cdot\mathbf{U}^*}{\Delta t}\biggr)\biggr|_{i,j}.
\end{equation}

The coefficients in the above equation are $A=\frac{1}{\rho_{i-1/2,j}\Delta x^2},~B=\frac{1}{\rho_{i+1/2,j}\Delta x^2},~D=\frac{1}{\rho_{i,j+1/2}\Delta y^2},~E=\frac{1}{\rho_{i,j-1/2}\Delta y^2}$ and $C=-\bigl(A+B+D+E\bigr)$. Given these coefficients, the pressure can be solved iteratively by using the following equation

\begin{equation}
\label{poi_it}
	p^{O+1}_{i,j} = \frac{1}{2}\biggl( 3p^{O+1}_{i,j} - p^O_{i,j} \biggr).
\end{equation}

In the above equation, $O$ denotes the iteration counter. Iteration of Eq. (\ref{poi2}) terminates until the absolute difference of the solutions obtained from two consecutive iterations becomes smaller than our chosen tolerance

\begin{equation}
	|p^{O+1}_{i,j} -p^{O}_{i,j} | < 10^{-5}.
\end{equation}

Substituting the computed pressure value $p^{n+1}$ into Eq. (\ref{vel_new}), the velocity field $\mathbf{U}^{n+1}$ is then obtained. It is noted that the continuity equation $\nabla\cdot\mathbf{U}^{n+1}$ is satisfied automatically in this method.

\subsection{Full solution algorithm}

\begin{figure}[p]
	\centering
	\begin{tikzpicture}[node distance = 2cm, auto]
	\node[ic](initial){Initialize $\phi^0,\mathbf{U}^0,p^0$};
	\node[ic, below of=initial](t_n){Loop starts, $\phi^n,\mathbf{U}^n,p^n$};
	\node[empty, below of= t_n](none1){};
	
	\node[solution, left of=none1, node distance =4cm](phi_n){$\phi^n=\hat{\phi^n}$};
	\node[solver, below of= phi_n](cLS){$\phi_t+\mathbf{U}\cdot\nabla\phi=0$};
	\node[solution, below of=cLS](cphi_new){$\phi^{n+1}$};
	\node[solver, below of= cphi_new](lam){$\lambda_I = \frac{\mathbb{M}(\phi^{0})-\mathbb{M}(\phi^{n+1})}{\Delta t\int_{\Omega}\rho(t)\delta^2(\hat\phi^n)|\nabla\hat\phi^n|~d\Omega}$};
	\node[solver, below of=lam](nLS){$\hat\phi_t+\mathbf{U}\cdot\nabla\hat\phi=\lambda_I\delta(\hat{\phi})|\nabla\hat{\phi}|$};
	\node[solution,below of=nLS, node distance=2cm](phi_new){$\hat{\phi}^{n+1*}$};
	\node[solver, below of=phi_new, node distance=2cm](reini){$\phi^*_\tau+\bar{S}(\phi^*_0)(|\nabla\phi^*|-1)=\lambda_R \delta(\phi^*)|\nabla\phi^*|$};
	\node[solution,below of=reini, node distance=2cm](phi_final){$\hat{\phi}^{n+1}$};
	
	\node[solver, right of=none1, node distance=4cm](rhomu){Calculate $\rho^n,\mu^n$ from Eq. (\ref{rhomu_norm}) };
	\node[solver, below of=rhomu, node distance=2.8cm](A){Calculate $\mathbf{A}^n =\mathbf{U}_t(\rho^n,\mu^n,\phi^n)-\rfrac{1}{\rho}\nabla p$ from Eq. (\ref{A})};
	\node[solver, below of=A, node distance=2.8cm](intermediate){Solve $\frac{\mathbf{U}^*-\mathbf{U}^n}{\Delta t} - (\rfrac{3}{2}\mathbf{A}^{n}-\rfrac{1}{2}\mathbf{A}^{n-1})=0$ for $\mathbf{U}^*$};
	\node[solver, below of=intermediate, node distance=2.8cm](rhomu2){Calculate $\rho^{n+1},\mu^{n+1}$ from Eq. (\ref{rhomu_norm})};
	\node[solver, below of=rhomu2, node distance=2.8cm](poi){Solve $\nabla\cdot(\frac{1}{\rho^{n+1}}\nabla p^{n+1})=\frac{1}{\Delta t}\nabla\cdot\mathbf{U}^*$ for $p^{n+1}$};
	\node[solver, below of=poi, node distance=2.8cm](vel){Find $\mathbf{U}^{n+1}$ by solving $\frac{\mathbf{U}^{n+1}-\mathbf{U}^*}{\Delta t} = - \frac{1}{\rho^{n+1}}\nabla p^{n+1}$};
	\node[ic, below of=t_n, node distance = 18cm](end){End of computation at $t=n\Delta t$};
	
	\path[line] (initial) -- (t_n);
	\path[line] (t_n) -| (phi_n);
	\path[line] (t_n) -| (rhomu);
	\path[line, dashed] (phi_n) -- (rhomu);
	\path[line] (phi_n) -- (cLS);
	\path[line](cLS)--(cphi_new);
	\path[line](cphi_new)--(lam);
	\path[line](lam)--(nLS);
	\path[line](phi_n)-|(-8,-12)--(nLS);
	\path[line](nLS)--(phi_new);
	\path[line](phi_new)--(reini);
	\path[line](reini)--(phi_final);
	\path[line](phi_final)|-(end);
	
	\path[line](rhomu)--(A);
	\path[line](A)--(intermediate);
	\path[line](intermediate)--(rhomu2);
	\path[line](rhomu2)--(poi);
	\path[line](poi)--(vel);
	\path[line](vel)|-(end);
	\path[line,dashed](phi_final)-| (-0.3,-12.4) -- (rhomu2);
	\path[line](end.south)|-(10,-21) |- ([yshift=1cm] t_n);
	\end{tikzpicture}
	\caption{Flow chart of the proposed solution algorithm.}
	\label{flowchart}
\end{figure}
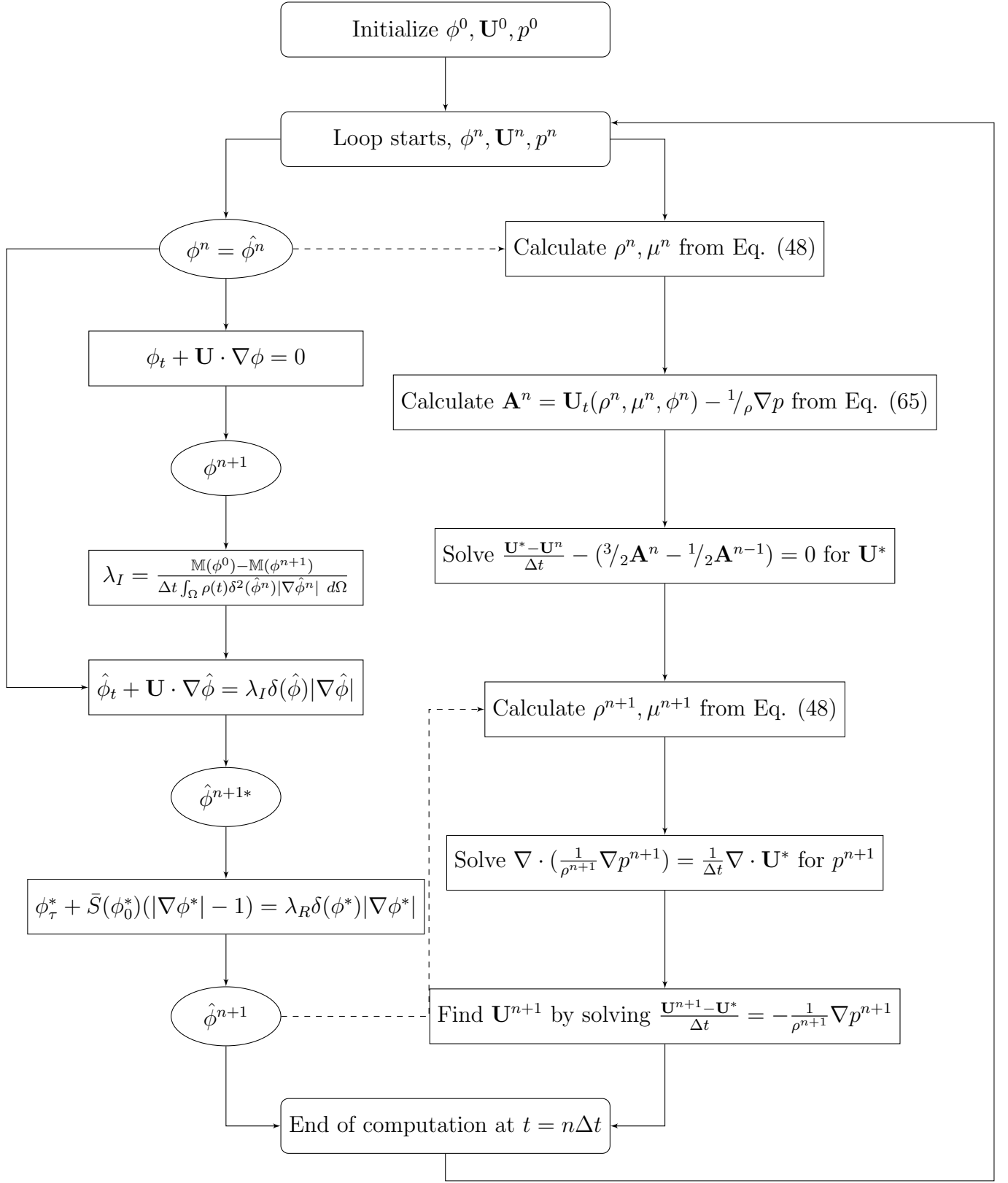

In summary, the motion of interface is captured by solving the Navier-Stokes equations given in Eqs. (\ref{continuity})-(\ref{momentum}) and our proposed mass-preserving governing equation for the level set function given in Eq. (\ref{mass-preserving ls}). Also, reinitialization of the level set value is performed to make the level set function to be a distance function. The solution algorithm is given in Fig. \ref{flowchart}.

\section{Validation studies}
To confirm the degree of mass conservation using the proposed mass-preserving level set method (\textbf{MPLS}) and the classical level set method (\textbf{LS}), two different error norms are introduced and defined as follows

\begin{align}
\label{errors}
\varepsilon_M(t) &= \frac{|M_0 - M(t)|}{M_0},\\
\bar{\varepsilon}_M &= \frac{1}{T}\int_0^T \varepsilon_M(t)~dt.
\end{align}

In the above equations, $M_0$ denotes the total mass of fluid of interest at $t=0$. To evaluate the stability of the solution obtained by different numerical schemes, we will calculate the values of $\varepsilon_M(t)$ for all $t$ that satisfies $T\ge t\ge0$. If $\varepsilon_M(t)$ grows as $t$ increases, one can say that the solution is unstable, and the solution will eventually blow up due to the discretization errors introduced into the simulation. Then, we will calculate the average loss of mass to show the loss of mass in the computation. Moreover, to show the computational efficiency of the \textbf{MPLS} method, we introduce the factor $\Theta^+$, which is expressed as follows:

\begin{equation}
	\Theta^+=\frac{\Theta_{MPLS}-\Theta_{LS}}{\Theta_{LS}}.
\end{equation} 
 
In the above equation, $\Theta_{MPLS}$ denotes the CPU time of the computation by using the \textbf{MPLS} method, while $\Theta_{LS}$ is the CPU time of computation by using the \textbf{LS} method. The value of $\Theta^+$ denotes the ratio of additional CPU time to implement the \textbf{MPLS} method in the simulation, which can be regarded as the level of sacrificing the computational efficiency. The time step $\Delta t$ and the mesh size $h$ are set as $\Delta t=0.1h$ in all problems described in this section.

\subsection{Two-dimensional vortex deforming problem}
The problem with a circle evolving with a prescribed velocity field was first studied by Rider and Kothe \cite{bib:Rider141(1998)}. In this problem, a circle with radius $r=0.15$ is initially located at $(x,y)=(0.5,0.75)$ in a square box $\Omega: [0,1]\times[0,1]$. The prescribed velocity is given as
\begin{align}
\label{vortex-velocity}
u(x,y,t) &= \sin^2(\pi x)\sin(2\pi y)\cos(\pi t/T),\\
v(x,y,t) &= -\sin(2\pi x)\sin^2(\pi y)\cos(\pi t/T).
\end{align}

The circle starts deforming its shape, and the circle will theoretically return back to its initial shape at $t=T$, which is set as $T=16$ in this study. During the shape deformation in the predicted solution, a very thin tail will be formed. The longer the period is, the thinner the filament will be. This problem has been considered as a standard comparison of different numerical methods.\\

\begin{figure}[p]
	\includegraphics[width=\textwidth]{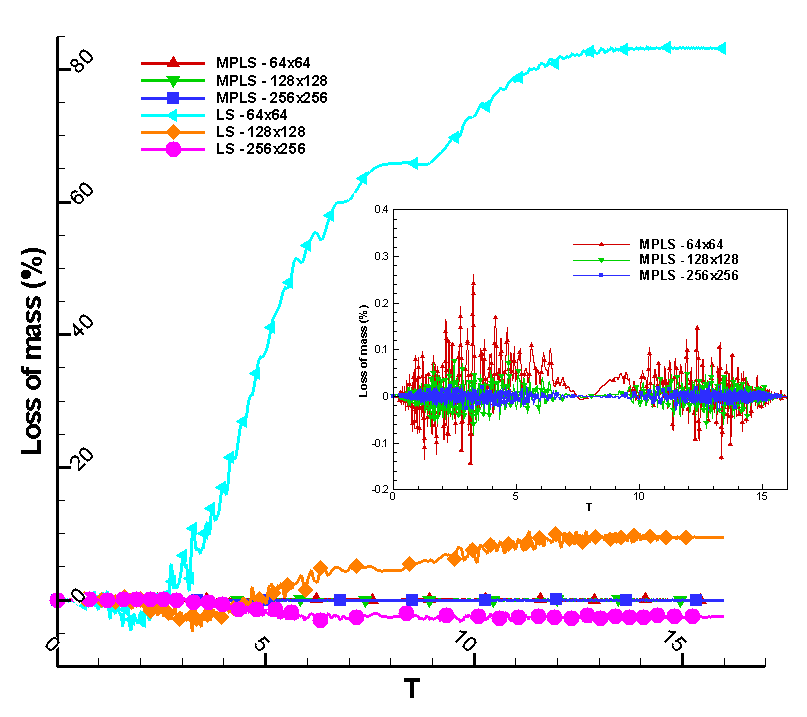}
	\caption{Comparison of the predicted percentages of the loss of mass using different methods with different grid numbers for the two-dimensional vortex deforming problem.}
	\label{vortex_mass}
\end{figure}

This problem will be simulated at three different grid numbers - $64^2, 128^2$ and $256^2$. The corresponding averaged loss of mass and CPU time are given in Table \ref{vortex2d}. According to these tabulated results, solutions obtained by the \textbf{MPLS} method have much smaller values of $\bar{\varepsilon}_M$ in comparison with that of the \textbf{LS} method. From the plot of loss of mass, cast in percentage form, shown in Fig. \ref{vortex_mass}, the value of $\varepsilon_M(t)$ obtained by the \textbf{LS} method increases much faster than that of the \textbf{MPLS} method. As a result, we can conclude that the  \textbf{MPLS} method is more efficient.\\

\begin{figure}
	
	\centering
	
	\begin{subfigure}[H]{0.6\textwidth}
		\includegraphics[width=\textwidth]{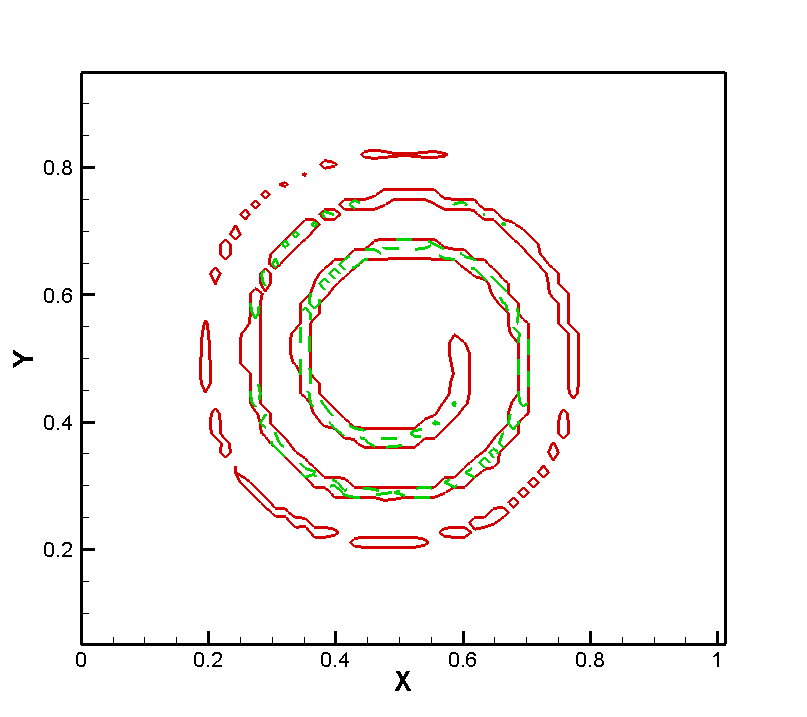}
		\caption{}
	\end{subfigure}

	\begin{subfigure}[H]{0.6\textwidth}
		\includegraphics[width=\textwidth]{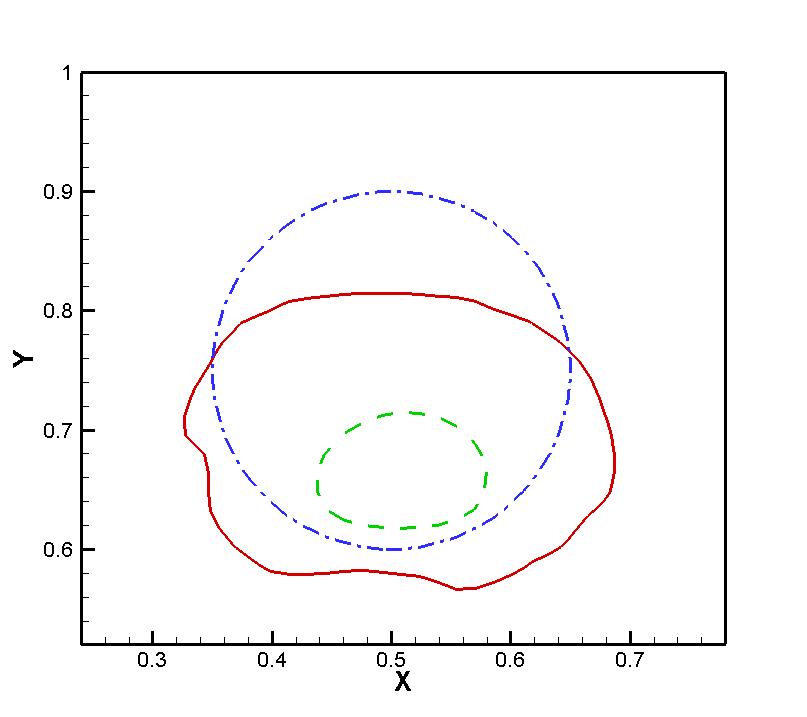}
		\caption{}
	\end{subfigure}	
	\caption{Comparison of the predicted interface for the two-dimensional vortex deforming problem in grids $64^2$. (a) $t=T/2=8$; (b) $t=T=16$. (Red solid denotes the solution obtained by \textbf{MPLS} method, green dash denotes the solution obtained by \textbf{LS} method, and blue dash-dot denotes the exact solution at $t=T$.) }
	\label{vortex2d_64}
\end{figure}

\begin{figure}
	
	\centering
	
	\begin{subfigure}[H]{0.6\textwidth}
		\includegraphics[width=\textwidth]{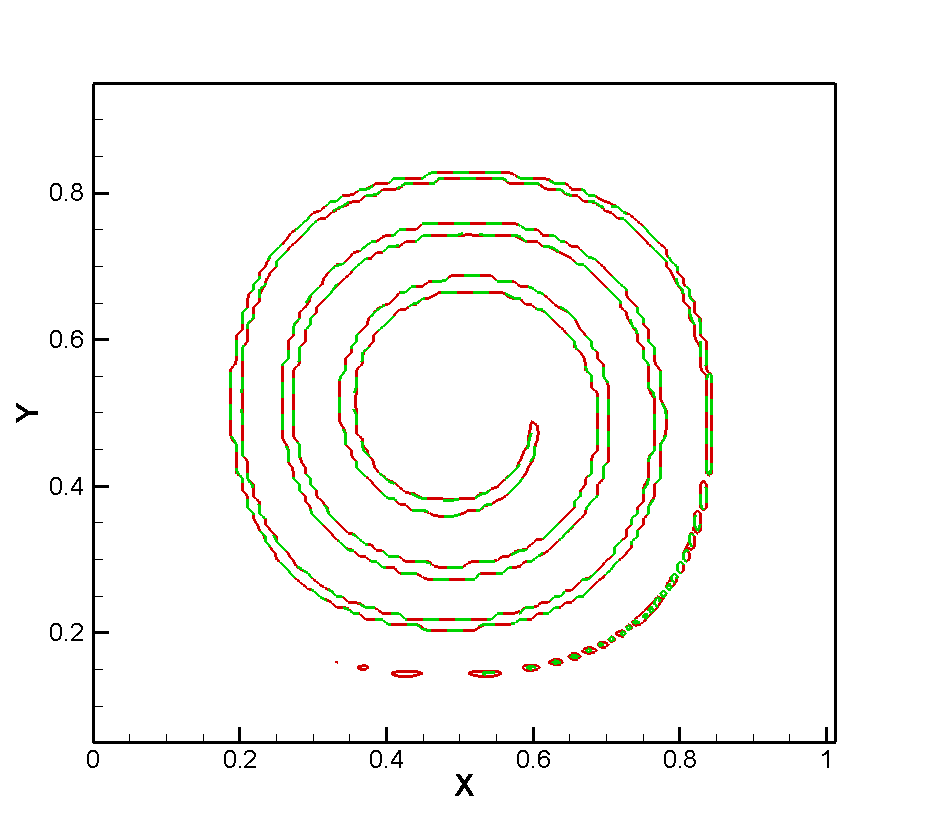}
		\caption{}
	\end{subfigure}
	
	\begin{subfigure}[H]{0.6\textwidth}
		\includegraphics[width=\textwidth]{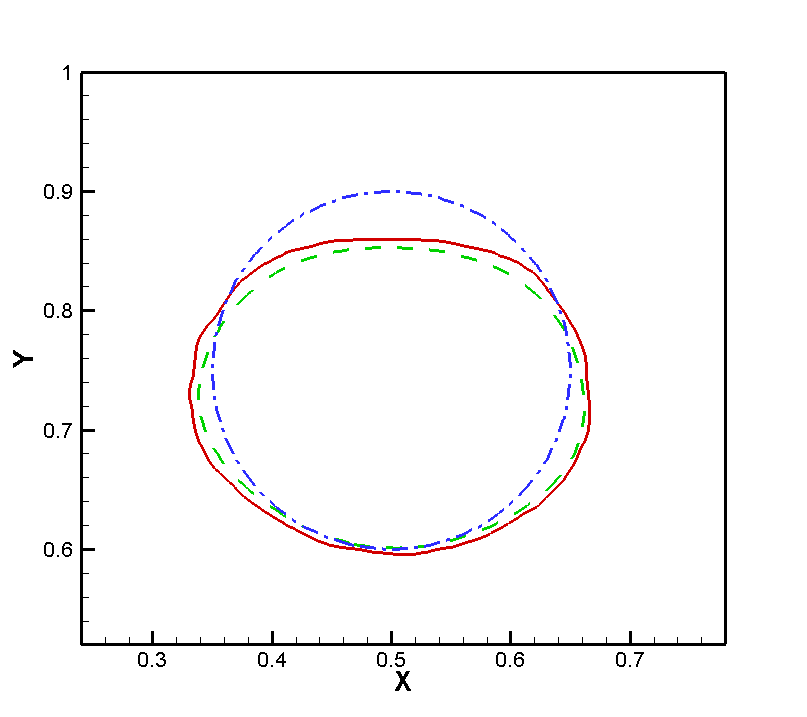}
		\caption{}
	\end{subfigure}	
	\caption{Comparison of the predicted interface for the two-dimensional vortex deforming problem in grids $128^2$. (a) $t=T/2=8$; (b) $t=T=16$. (Red solid denotes the solution obtained by \textbf{MPLS} method, green dash denotes the solution obtained by \textbf{LS} method, and blue dash-dot denotes the exact solution at $t=T$.) }
	\label{vortex2d_128}
\end{figure}

\begin{figure}
	
	\centering
	
	\begin{subfigure}[H]{0.6\textwidth}
		\includegraphics[width=\textwidth]{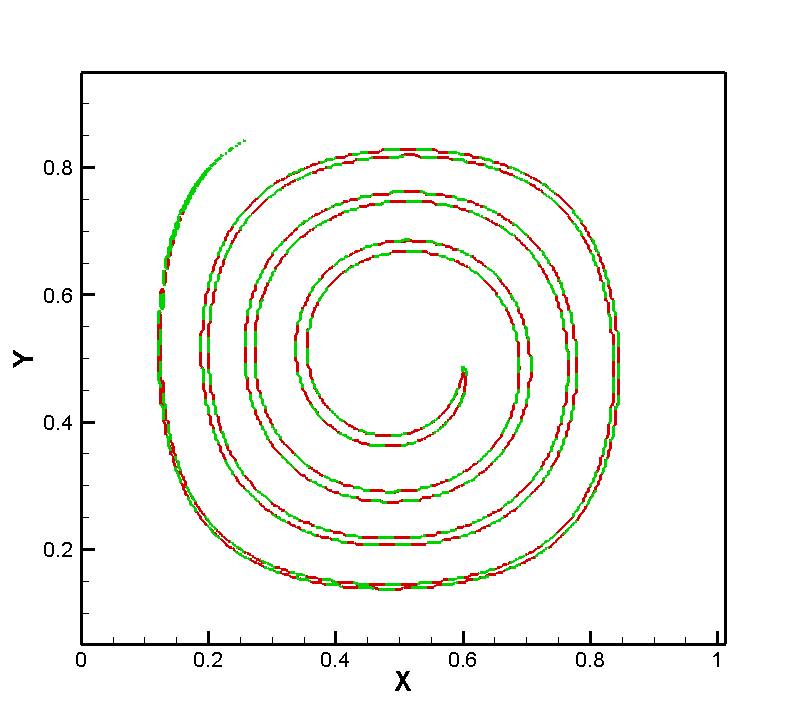}
		\caption{}
	\end{subfigure}
	
	\begin{subfigure}[H]{0.6\textwidth}
		\includegraphics[width=\textwidth]{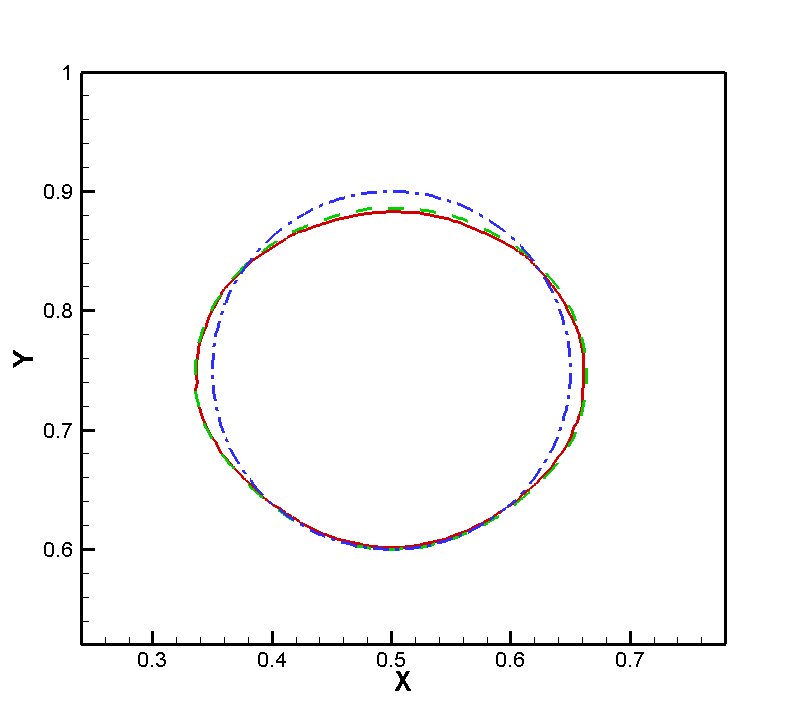}
		\caption{}
	\end{subfigure}	
	\caption{Comparison of the predicted interface for the two-dimensional vortex deforming problem in grids $256^2$. (a) $t=T/2=8$; (b) $t=T=16$. (Red solid denotes the solution obtained by \textbf{MPLS} method, green dash denotes the solution obtained by \textbf{LS} method, and blue dash-dot denotes the exact solution at $t=T$.) }
	\label{vortex2d_256}
\end{figure}

Owing to the great ability of the \textbf{MPLS} method to retain mass, capturing of the thin tail - obtained in the domain with the grid number $64^2$ - can be seen in such a coarse mesh as shown in Fig. \ref{vortex2d_64}. However, the solution obtained by the \textbf{LS} method is strongly affected by its poor conservation of mass, in the sense that the thin tail has been smeared a lot. In the fine grid simulation, solutions obtained by the \textbf{MPLS} and \textbf{LS} methods both have good agreement with the exact solution as shown in Figs. \ref{vortex2d_128}-\ref{vortex2d_256}. Regarding the issue of computational efficiency, roughly an additional 10\% of  CPU time is required to get the solutions when employing the \textbf{MPLS} method.

\subsection{Three-dimensional vortex deforming problem}
In order to show the scheme ability of retaining mass conservation in three dimensional simulation, single vortex deforming problem will be simulated using the proposed method and the pure level set method. This problem was first introduced by LeVeque \cite{bib:LeVeque(1996)}. A sphere of radius $r=0.15$ is located at $(x,y,z)=(0.35,0.35,0.35)$ in a cubic domain $\Omega:[0,1]\times[0,1]\times[0,1]$. Velocity components considered in this problem at any time $0\le t\le T$ are given by

\begin{equation}
\label{vortex3d-vel}
\begin{aligned}
u(x,y,z,t) &= 2\sin^2(\pi x)\sin(2\pi y)\sin(2\pi z)\cos(\pi t/T), \\
v(x,y,z,t) &= -\sin(2\pi x)\sin^2(\pi y)\sin(2\pi z)\cos(\pi t/T), \\
w(x,y,z,t) &= -\sin(2\pi x)\sin(2\pi y)\sin^2(\pi z)\cos(\pi t/T).
\end{aligned}
\end{equation}

\begin{figure}[p]
	\includegraphics[width=\textwidth]{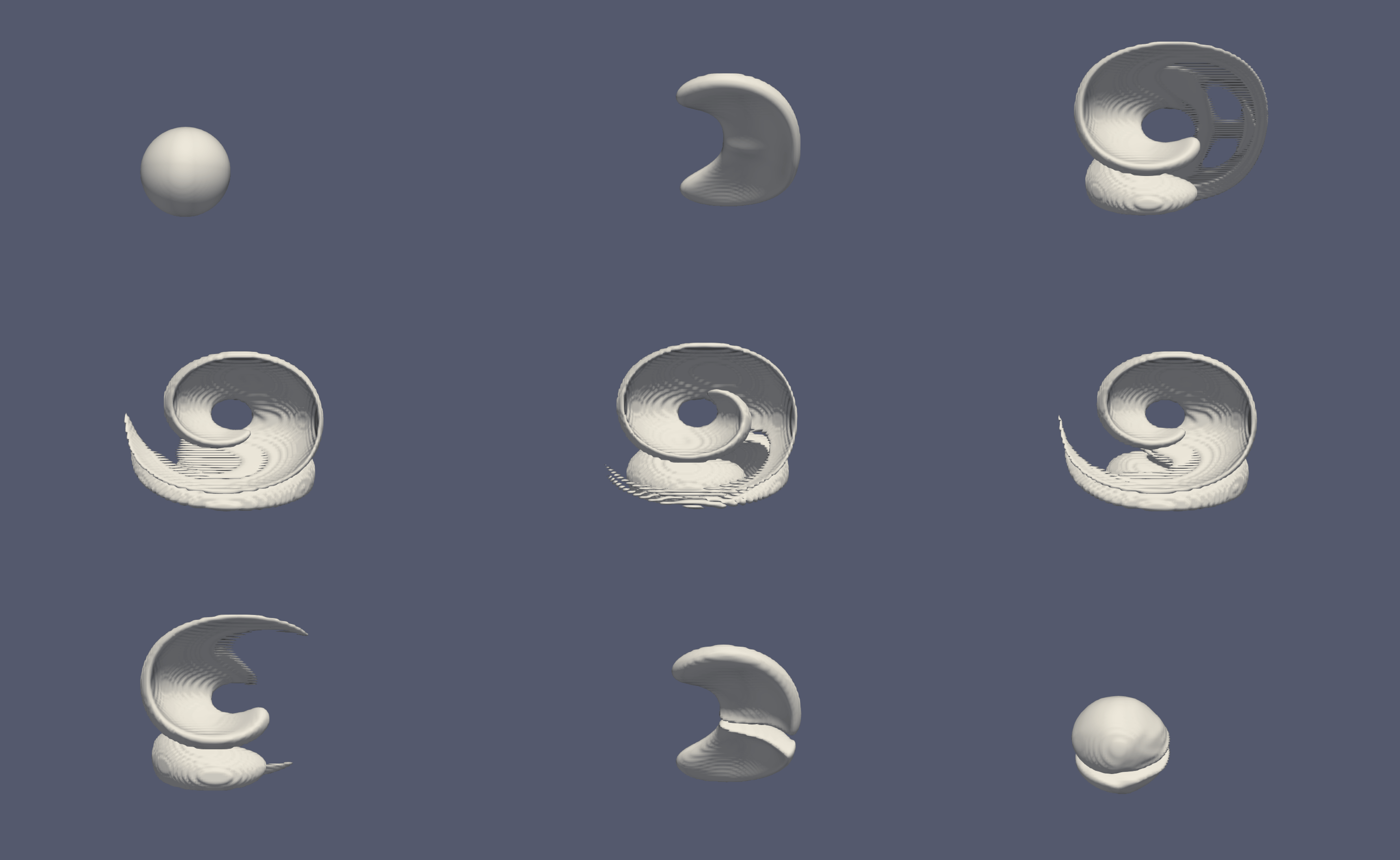}
	\caption{Snapshots of the predicted interfaces of the three-dimensional vortex deforming predicted in grids $128^3$. From top to down, from left to right, $t=0.0, 0.5, 1.0, 2.0, 3.0, 4.0, 5.0, 5.5, 6.0$.}
	\label{vortex3d}
\end{figure}

\begin{figure}
	
	\centering
	
	\begin{subfigure}[H]{0.6\textwidth}
		\includegraphics[width=\textwidth]{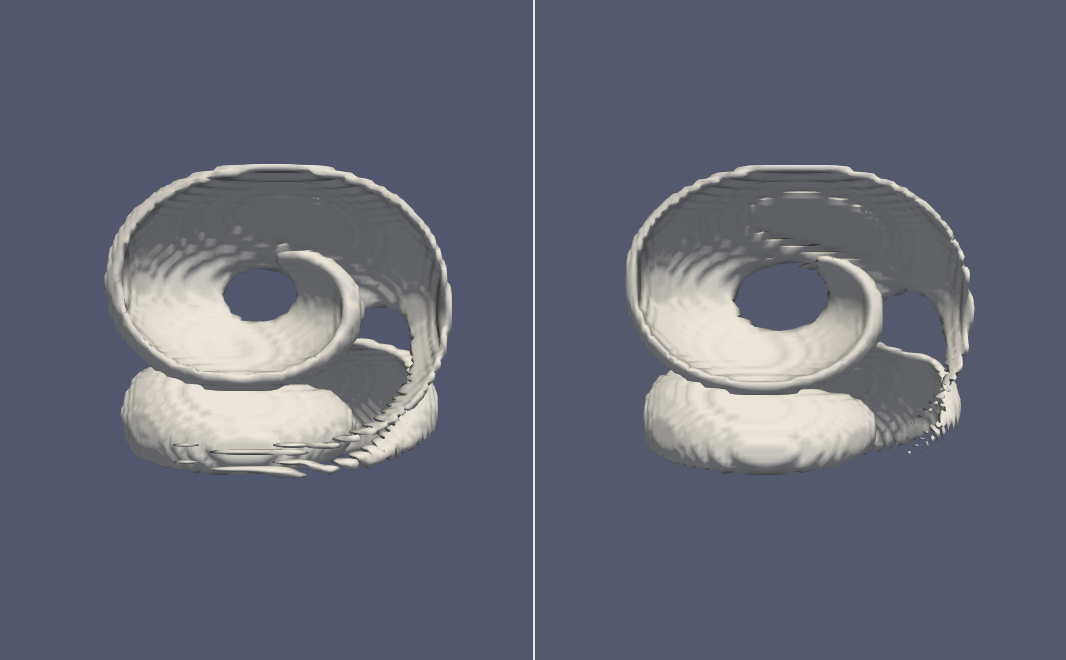}
		\caption{}
	\end{subfigure}
	
	\begin{subfigure}[H]{0.6\textwidth}
		\includegraphics[width=\textwidth]{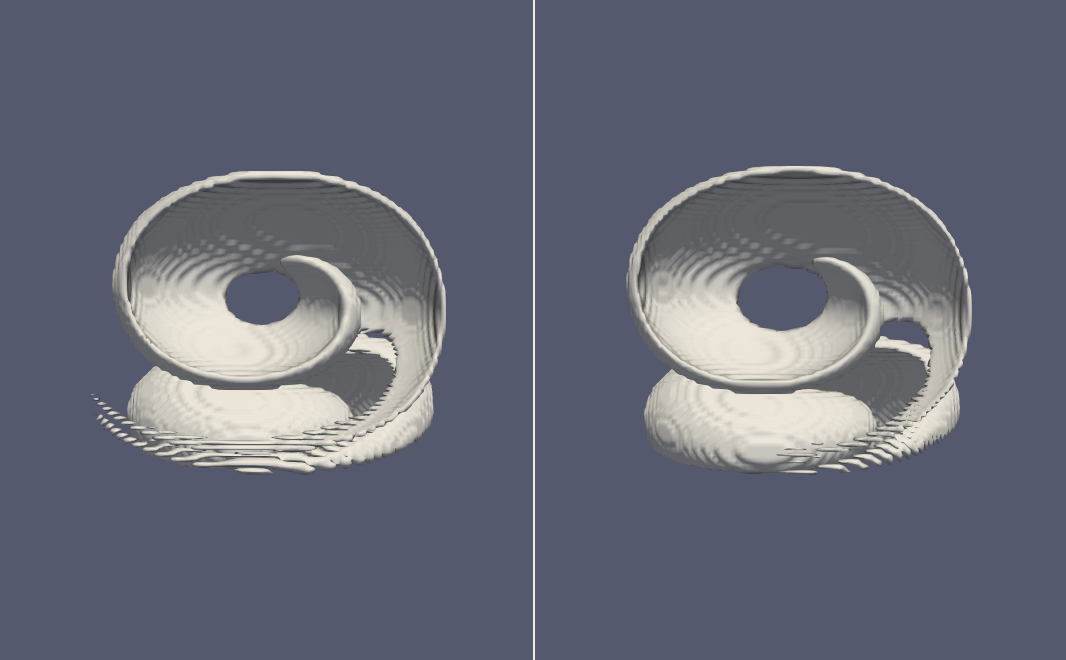}
		\caption{}
	\end{subfigure}	
	\caption{Comparison of the predicted interface for the three-dimensional vortex deforming problem at $t=3.0$. (a) In grid number $96^3$; (b) In grid number $128^3$. (The left-hand side of each figure denotes the solution obtained by \textbf{MPLS} method, and the right-hand side denotes the solution obtained by \textbf{LS} method) }
	\label{vortex3d_compare}
\end{figure}

\begin{figure}[p]
	\includegraphics[width=\textwidth]{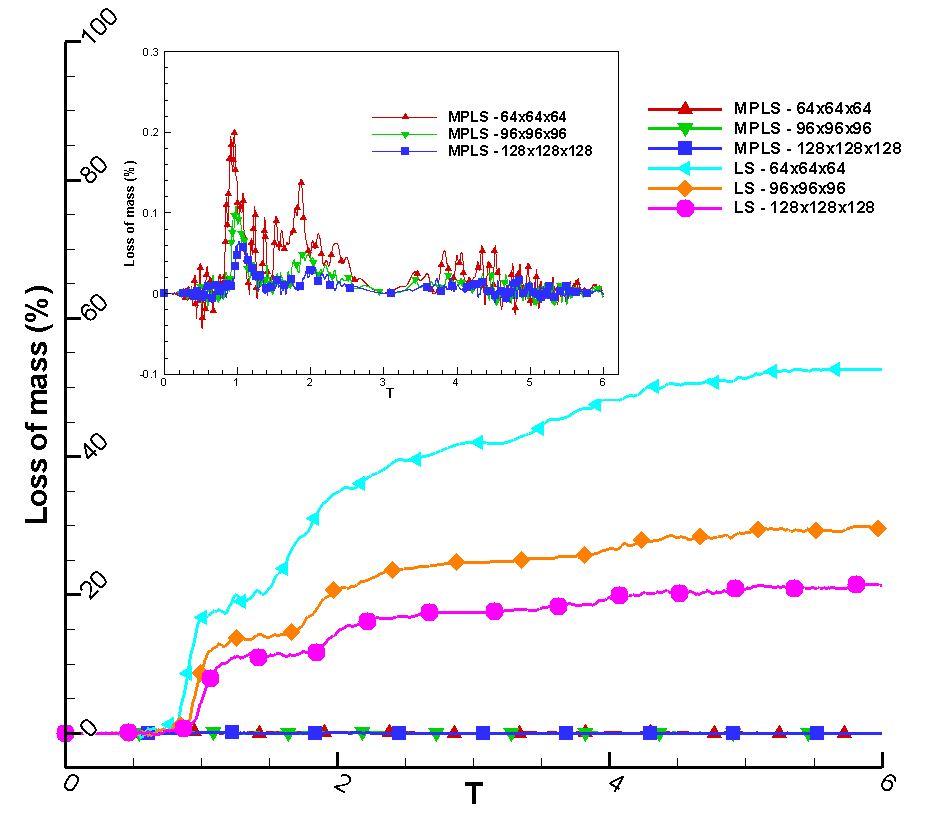}
	\caption{Comparison of the predicted percentages of the loss of mass using different methods with different grid numbers for the three-dimensional vortex deforming problem.}
	\label{vortex3d_mass}
\end{figure}

Note that $T$ is the period of the rotating shear vortex. The sphere shall be reversed back to its initial shape at $t=T$, which is set as $6.0$ in this study. Eq. (\ref{vortex3d-vel}) allows the sphere to start deforming, and, then, evolves to form two vortices that scoop out the opposite side of the sphere. This problem has been simulated in three different grids with $64^3, 96^3$ and $128^3$ nodal points. The evolution of the vortex using the grid $128^3$ is given in Fig. \ref{vortex3d}.  In Fig. \ref{vortex3d_compare} , the solution at $t=3.0$ obtained by using the \textbf{MPLS} and \textbf{LS} methods are used to show the great ability of retaining mass using the \textbf{MPLS} method. One can see that a very thin film is formed at $t=3.0$, and it can be clearly seen in the solutions obtained by the \textbf{MPLS} method, while this thin film is considerably smeared in the solutions obtained by the \textbf{LS} method. Moreover, the average mass loss $\bar\varepsilon_M$ obtained in different grids are given in Table \ref{vortex3d_table}. According to the tabulated results, the value of $\bar\varepsilon_M$ obtained by using the \textbf{MPLS} method is much smaller than that using the \textbf{LS} method. It is noted that 12\% of additional CPU time is needed to implement the \textbf{MPLS} method. Also, the method of \textbf{MPLS} is more stable than the method of \textbf{LS}, since the value of mass loss does not grow significantly, as shown in Fig. \ref{vortex3d_mass}. 

\section{Numerical results}
Droplet collision and impact are omnipresent phenomena and are well known to occur in nature and in processing industries. A profound understanding and a better control of the subsequent events after impact are crucial to many applications. For instance, spreading is desirable for coating or ink-jet printing while splashing may improve the efficiency of evaporation and mixing in fuel combustion \cite{bib:Josserand(2016), bib:Agbaglah(2015)}. The outcome of the impact  depends on various factors, including the speed and the type of fluids, and on the substrate. If the substrate is dry, results will depend on the wetting ability and the smoothness of the surface. Droplets impacting into a deep pool or thin layer lead to another degree of physical complexity, as evidenced by the presence of a dramatic change in topology resulting from the surface tension and capillary instability. A thorough understanding of droplets impact into a liquid surface is still lacking. The subjects of these unexplored complex dynamics about droplet impact include, for example, the understanding of formation of singular surface deformation and the accompanying flow instabilities of different physical kinds and the transition from splashing to spreading \cite{bib:Josserand(2016)}. However, relation among all the selected dimensionless parameters and the resulting outcomes have been discussed by many researches \cite{bib:Riboo(2003), bib:Wang(2000)}.

\subsection{A single droplet impact on a liquid pool}
Here, we investigate the falling of a liquid drop onto a liquid surface and compare our simulations with the experimental results of Wang et al. \cite{bib:Wang(2013)}. We set our computational domain as $\Omega:[0,5]\times[0,5]\times[0,7]$. The liquid droplet of diameter $5.65mm$ is initialized just above the liquid pool ($\approx 0.1$) with the initial velocity $0.953m/s$, where the depth of the pool is set at $4.0D$ ($D$ is the diameter of droplet). The parameters of the problem are identical to the case \textbf{L1} in the paper of A.-B. Wang et al. \cite{bib:Wang(2013)}. We set the Reynolds numbers as $Re=4790$, the Weber number as $We=77$, and Froude number as $Fr=4.12$. The air water density ratio and the viscosity ratio are chosen to be $\rho_{g}/\rho_{l}=0.001$ and $\mu_{g}/\mu_{l}=0.001$. \\

\begin{figure}
	
	\centering
	
	\begin{subfigure}[H]{0.6\textwidth}
		\includegraphics[width=\textwidth]{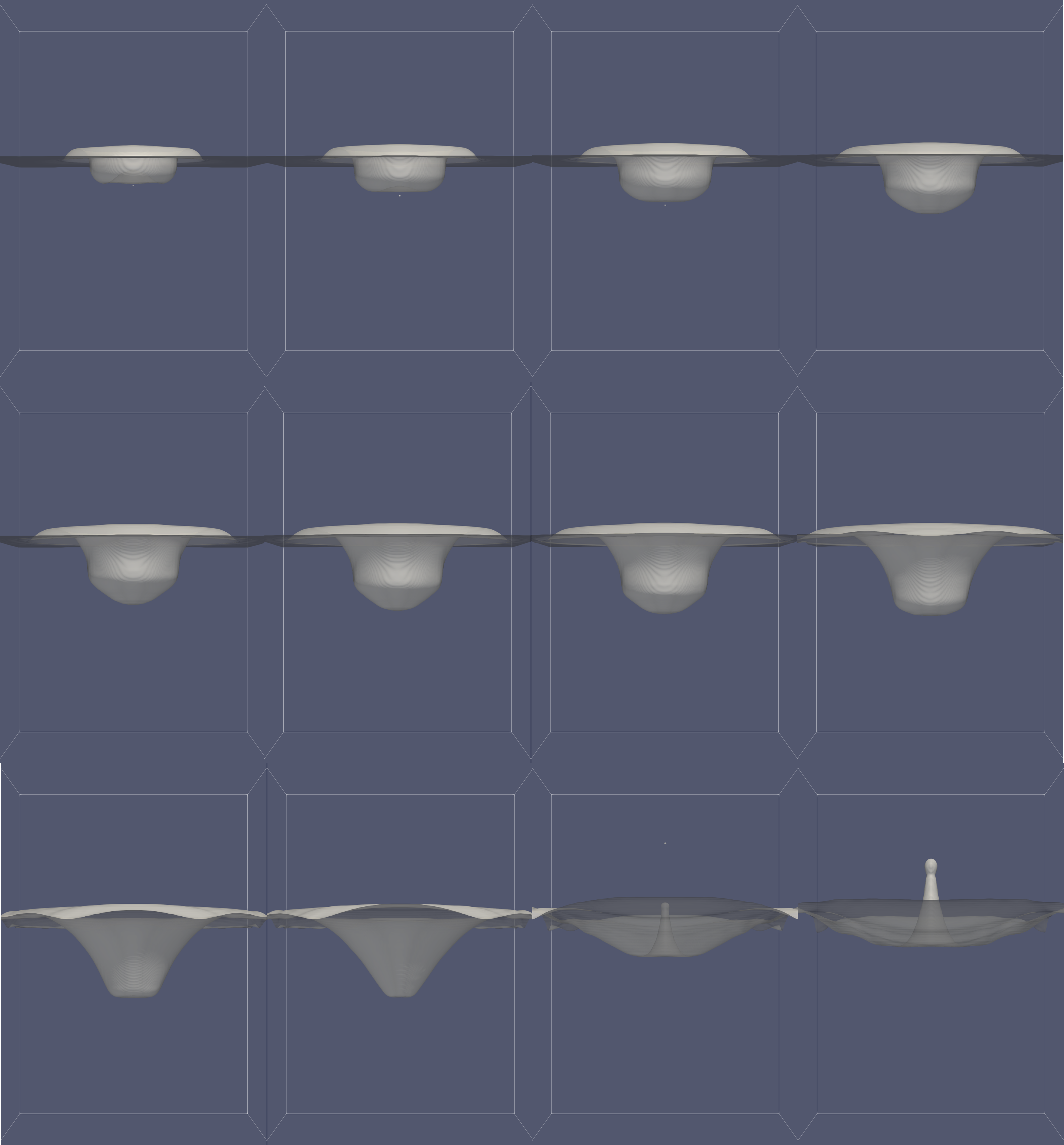}
		\caption{}
	\end{subfigure}
	
	\begin{subfigure}[H]{0.6\textwidth}
		\includegraphics[width=\textwidth]{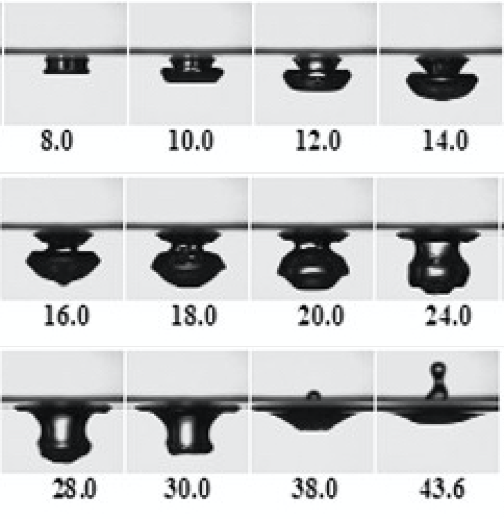}
		\caption{}
	\end{subfigure}	
	\caption{Comparison of the predicted interfaces using the \textbf{MPLS}  method in grids $200\times200\times280$ for the problem of droplet impact into a deep pool. (a) Present; (b) An-Bang Wang et. al \cite{bib:Wang(2013)}.}
	\label{di}
\end{figure}

\begin{figure}[p]
	\includegraphics[width=\textwidth]{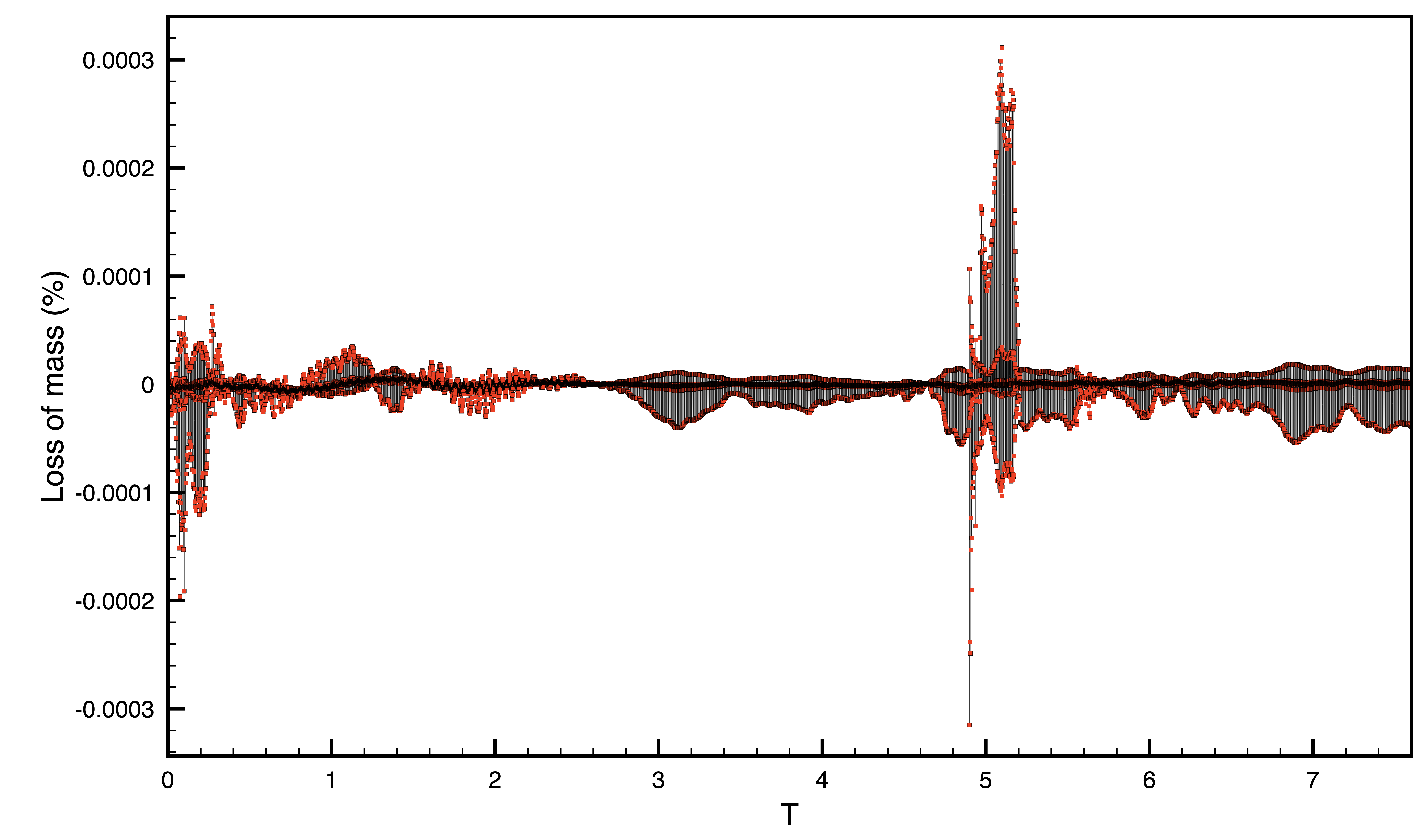}
	\caption{Predicted percentages of the loss of mass using the \textbf{MPLS} method in grids $200\times200\times280$ for the problem of droplet impact into a deep pool.}
	\label{di_mass}
\end{figure}

Snapshots of the interfaces predicted in the grid of $200\times200\times280$ points are given in Fig. \ref{di}, and the evolution of mass loss, cast in percentage form, are displayed in Fig. \ref{di_mass}. According to Fig. \ref{di}, we have seen a good match with the experimental results. The phenomenon of jet is seen to be well captured in our simulations. Moreover, the discrepancy between the predicted and theoretical mass is only $0.0003\%$ in our simulation.

\subsection{A single drop impact on a liquid layer}

This problem has been investigated by many research groups to justify their simulation ability of capturing a complicated topology change, both in two-dimensional \cite{bib:Harlow(1967)} and three-dimensional \cite{bib:Yokoi(2008),bib:Xiao(2005)m} simulations. The initial setup in this study is identical to that of in the paper of Kensuke Yokoi \cite{bib:Yokoi(2008)}. A droplet of diameter $5.33mm$ with the initial velocity $2.0m/s$ is impacting into a liquid film of $1mm$ depth. We set the Reynolds number as $Re=6270.58$, the Weber number as $We=426.4$, and the Froude number as $Fr=8.75$. The air water density ratio and the viscosity ratio are chosen to be $\rho_{g}/\rho_{l}=0.0013$ and $\mu_{g}/\mu_{l}=0.0006$. The computational domain is set in $\Omega:[-3,3]\times[-3,3]\times[0,3]$.\\

\begin{figure}[p]
	\includegraphics[width=\textwidth]{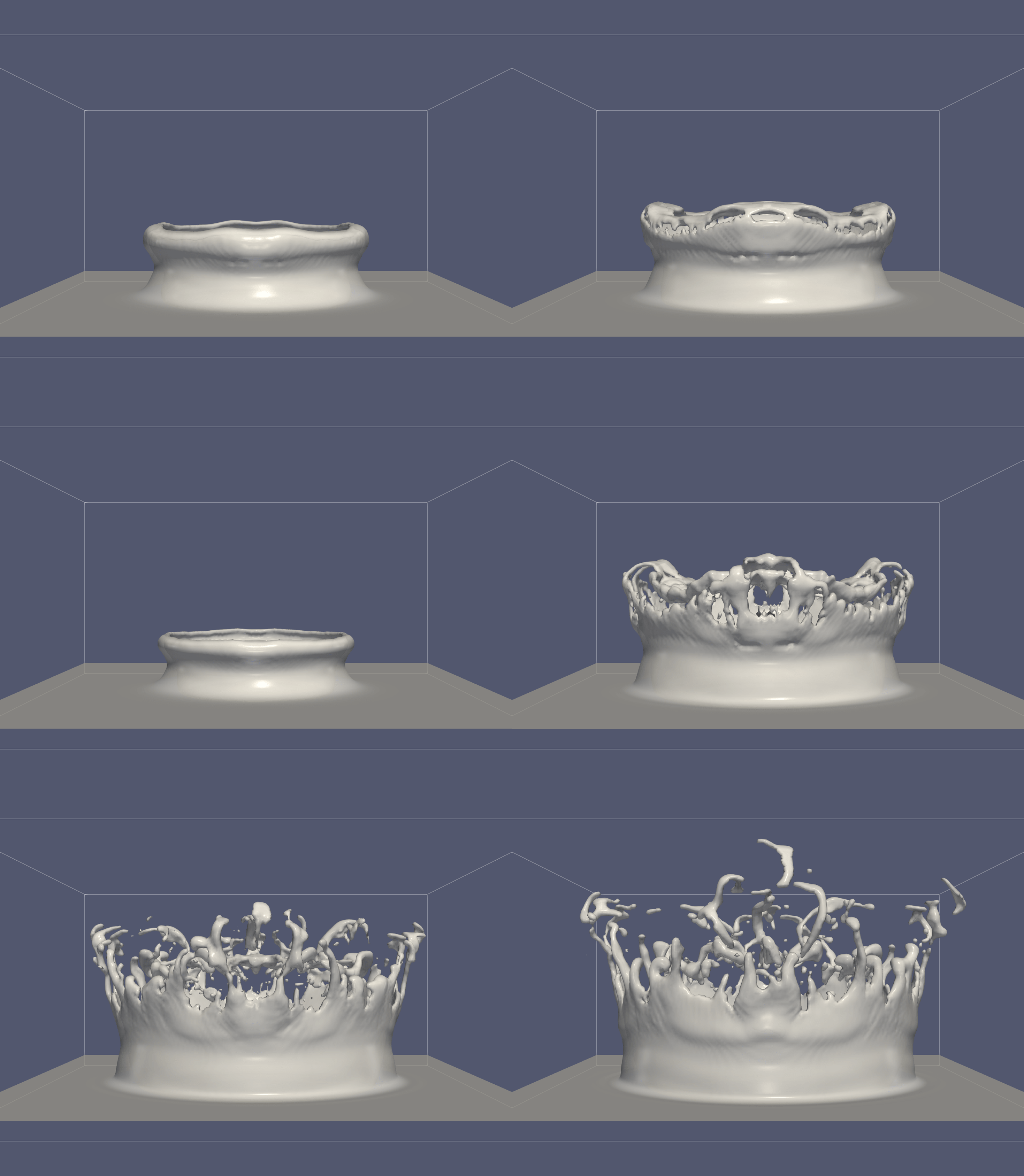}
	\caption{Predicted interfaces using the \textbf{MPLS} method in grids $210\times210\times105$ for the problem of droplet impacting upon a thin liquid layer.}
	\label{mc}
\end{figure}

\begin{figure}[p]
	\includegraphics[width=\textwidth]{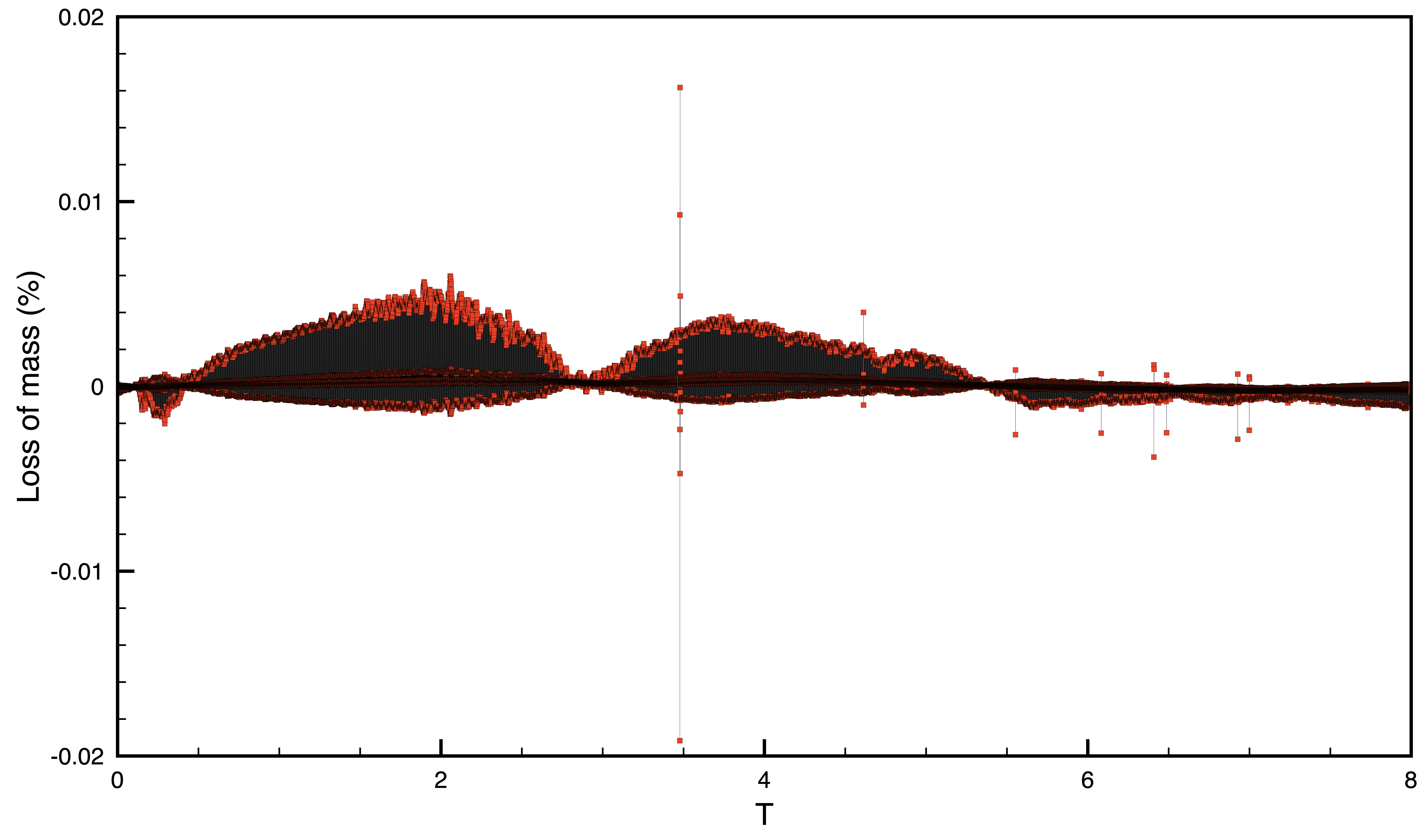}
	\caption{Predicted percentages of the loss of mass using the \textbf{MPLS} method in grids $210\times210\times105$ for the problem of droplet impacting upon a thin liquid layer.}
	\label{mc_mass}
\end{figure}

Snapshots of the predicted interfaces in $200\times200\times280$ are shown in Fig. \ref{mc}. The corresponding loss of mass is also depicted in Fig. \ref{mc_mass}. According to Fig. \ref{mc}, our simulated results are seen to be similar to those predicted by Kensuke Yokoi  in his paper \cite{bib:Yokoi(2008)}. According to Fig. \ref{mc_mass}, the percentage of the predicted error in mass is about $0.01\%$. As a result, the quality of the solutions is verified from the numerical point view.

\section{Concluding remarks}
In this paper, a mass-preserving level set method has been developed to capture evolving interfaces. An additional source term is added to the original level set method to reduce the discretization error that leads to the imbalance of mass.\\

We implement the proposed mass-preserving level set method to four different verification studies  in two and three dimensions. According to the tabulated results, solutions solved by using the mass-preserving level set method can retain its mass very well without requiring a significant amount of CPU time (less than $13\%$).\\

Mass-preserving level set method has been applied to practical two-phase flows as well, including a single droplet impacting on a deep liquid pool and a thin liquid layer. In our simulations, solutions obtained by using the mass-preserving method can preserve its mass very well after the dramatic topology change. As a result, it is reasonable and cost-effective to apply our proposed mass-preserving level set method to predict two-phase flows.

\begin{table}[p]
	\centering
	\caption{Comparison of the average mass loss and CPU time for the two-dimensional vortex deforming problem at different grid numbers. }
	\begin{tabular}{p{0.2\textwidth}p{0.25\textwidth}p{0.25\textwidth}p{0.25\textwidth}}
		\hline\\[-0.4cm]
		\multicolumn{4}{c}{Grid number} \\
		 & $64^2$ & $128^2$ & $256^2$ \\
		\hline\hline\\
		 \multicolumn{4}{l}{\textbf{MPLS}} \\
		 
	     $\bar\varepsilon_M~(\rho_{12}=1)$ & 
	     \s{1.1249}{-13} & \s{7.7902}{-15} & \s{9.4510}{-16}\\
	     
	     $\bar\varepsilon_M~(\rho_{12}=0.1)$ & 
	     \s{6.7474}{-11} & \s{8.4374}{-12} & \s{1.4145}{-12}\\
	     
	     $\bar\varepsilon_M~(\rho_{12}=0.01)$ & 
	     \s{9.1269}{-11} & \s{1.1651}{-11} & \s{1.9067}{-12}\\

	     CPU time(s) & 7.04 & 25.08 & 155.09\\[0.5cm]
		 \multicolumn{4}{l}{\textbf{LS}} \\
		 
		 $\bar\varepsilon_M~(\rho_{12}=1)$ & 
	     \s{3.8132}{-1} & \s{4.1193}{-2} & \s{8.4670}{-3}\\
	     
	     $\bar\varepsilon_M~(\rho_{12}=0.1)$ & 
	     \s{4.0708}{-1} & \s{4.4356}{-2} & \s{8.5046}{-3}\\
	     
	     $\bar\varepsilon_M~(\rho_{12}=0.01)$ & 
	     \s{4.0984}{-1} & \s{4.4687}{-2} & \s{8.5105}{-3}\\

	     CPU time(s) & 6.09 & 23.67 & 149.99\\[0.5cm]
	  	 $\Theta^+$& 15.59\% & 5.95\% & 3.40\%\\
		\hline
		\label{vortex2d}
	\end{tabular}
\end{table}

\begin{table}[p]
	\centering
	\caption{Comparison of the average mass loss and CPU time for the three-dimensional vortex deforming problem at different grid numbers. }
	\begin{tabular}{p{0.2\textwidth}p{0.25\textwidth}p{0.25\textwidth}p{0.25\textwidth}}
		\hline\\[-0.4cm]
		\multicolumn{4}{c}{Grid number} \\
		 & $64^3$ & $128^3$ & $256^3$ \\
		\hline\hline\\
		 \multicolumn{4}{l}{\textbf{MPLS}} \\
		 
	     $\bar\varepsilon_M~(\rho_{12}=1)$ & 
	     \s{1.8621}{-13} & \s{1.5031}{-13} & \s{6.5469}{-15}\\
	     
	     $\bar\varepsilon_M~(\rho_{12}=0.1)$ & 
	     \s{5.1781}{-11} & \s{3.5542}{-12} & \s{5.7381}{-13}\\
	     
	     $\bar\varepsilon_M~(\rho_{12}=0.01)$ & 
	     \s{9.1269}{-11} & \s{1.1651}{-11} & \s{1.9067}{-12}\\

	     CPU time(s) & 211.96 & 2824.94 & 41513.99\\[0.5cm]
		\multicolumn{4}{l}{\textbf{LS}} \\
		 
	     $\bar\varepsilon_M~(\rho_{12}=1)$ & 
	     \s{6.7649}{-1} & \s{3.0018}{-1} & \s{1.2490}{-1}\\
	     
	     $\bar\varepsilon_M~(\rho_{12}=0.1)$ & 
	     \s{7.0478}{-1} & \s{3.1022}{-1} & \s{1.2678}{-1}\\
	     
	     $\bar\varepsilon_M~(\rho_{12}=0.01)$ & 
	     \s{7.0791}{-1} & \s{3.1128}{-1} & \s{1.2697}{-1} \\

	     CPU time(s) & 197.10 & 2609.77 & 37192.29 \\[0.5cm]
		$\Theta^+$& 7.63\% & 8.24\% & 11.62\%\\
		\hline
		\label{vortex3d_table}
	\end{tabular}
\end{table}

\begin{figure}[p]
	\includegraphics[width=\textwidth]{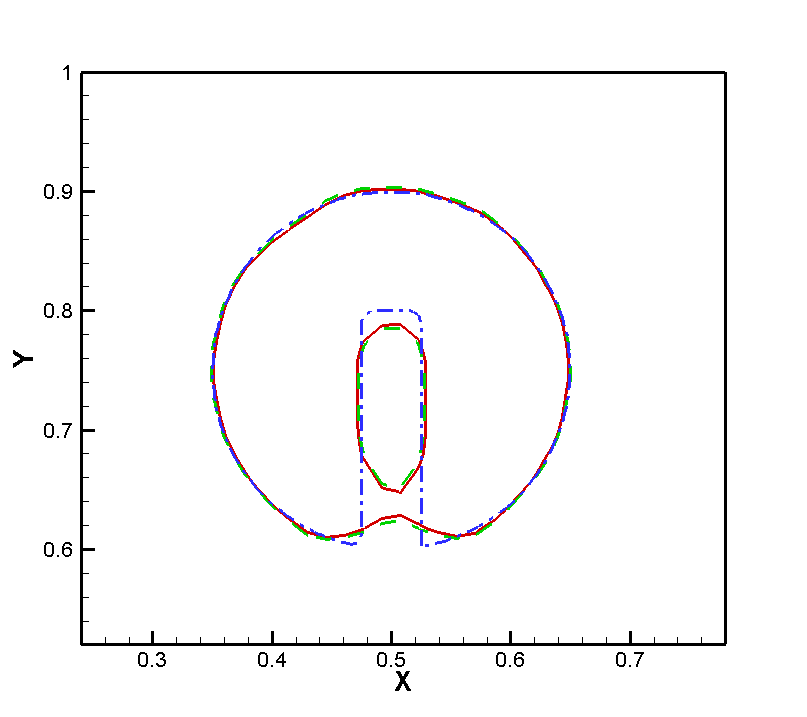}
	\caption{Comparison of the predicted interface for the two-dimensional rotating disk problem after ten revolutions in grids $64^2$. (Red solid denotes the solution obtained by \textbf{MPLS} method, green dash denotes the solution obtained by \textbf{LS} method, and blue dash-dot denotes the exact solution.)}
	\label{disk64}
\end{figure}

\begin{figure}[p]
	\includegraphics[width=\textwidth]{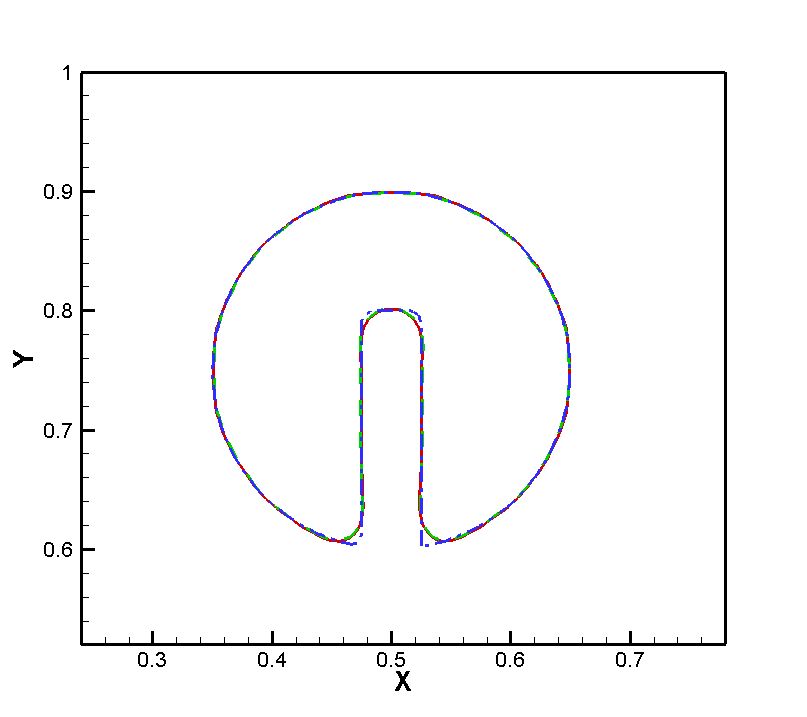}
	\caption{Comparison of the predicted interface for the two-dimensional rotating disk problem after ten revolutions in grids $128^2$. (Red solid denotes the solution obtained by \textbf{MPLS} method, green dash denotes the solution obtained by \textbf{LS} method, and blue dash-dot denotes the exact solution.)}
	\label{disk128}
\end{figure}

\begin{figure}[p]
	\includegraphics[width=\textwidth]{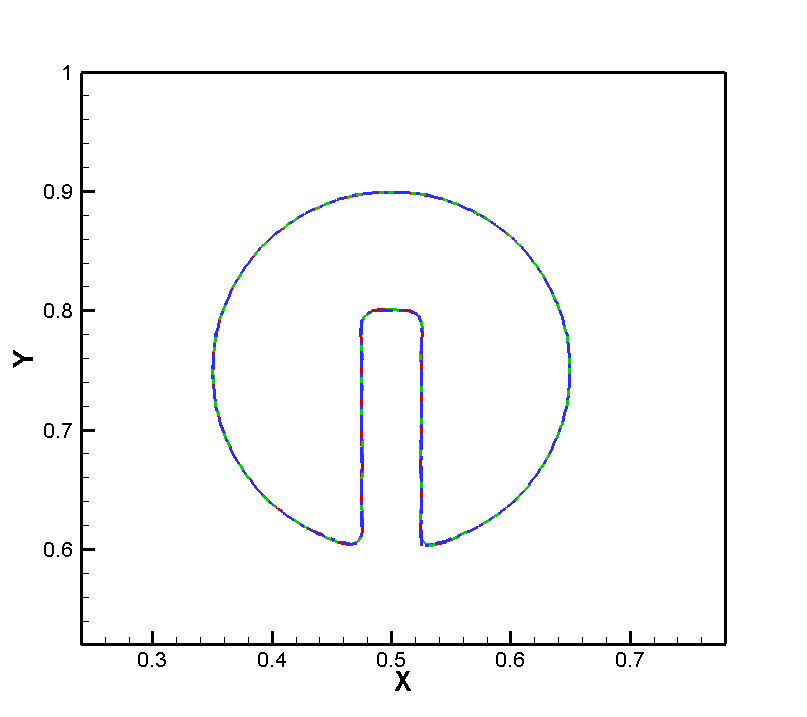}
	\caption{Comparison of the predicted interface for the two-dimensional rotating disk problem after ten revolutions in grids $256^2$. (Red solid denotes the solution obtained by \textbf{MPLS} method, green dash denotes the solution obtained by \textbf{LS} method, and blue dash-dot denotes the exact solution.)}
	\label{disk256}
\end{figure}

\begin{figure}[p]
	\includegraphics[width=\textwidth]{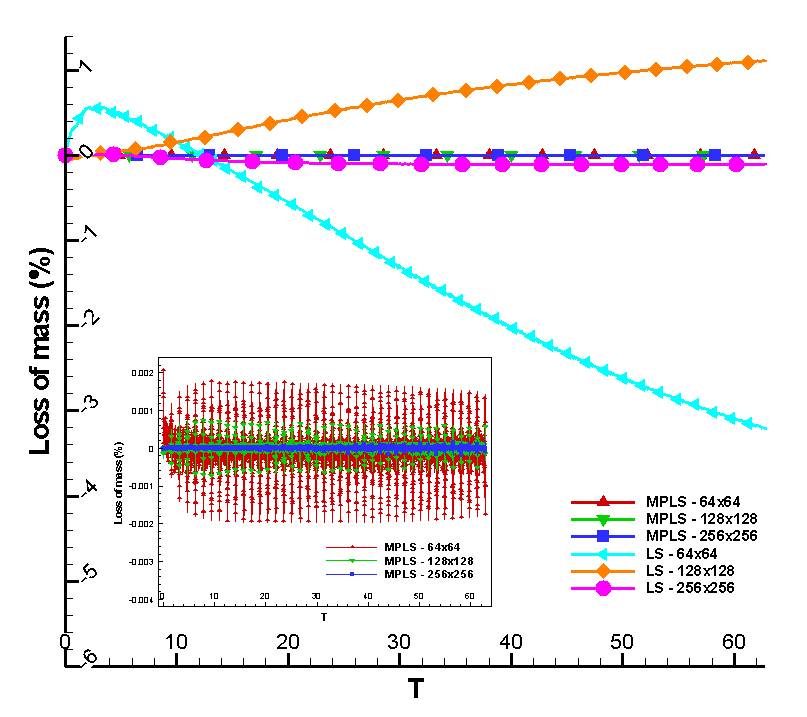}
	\caption{Comparison of the predicted percentages of the loss of mass using different methods with different grid numbers for the two-dimensional rotating disk problem.}
	\label{disk_mass}
\end{figure}

\begin{figure}
	
	\centering
	
	\begin{subfigure}[H]{0.6\textwidth}
		\includegraphics[width=\textwidth]{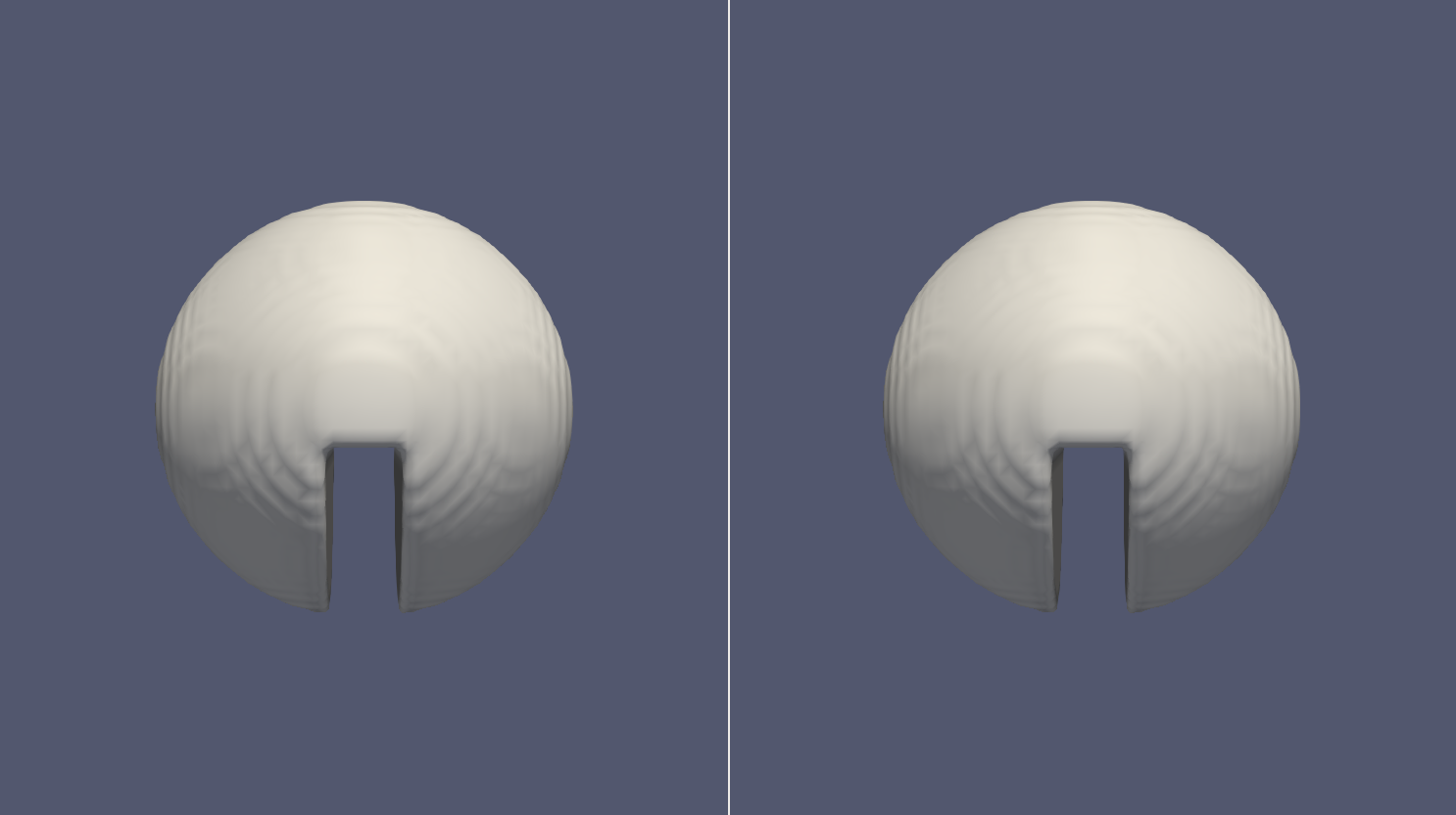}
		\caption{}
	\end{subfigure}
	
	\begin{subfigure}[H]{0.6\textwidth}
		\includegraphics[width=\textwidth]{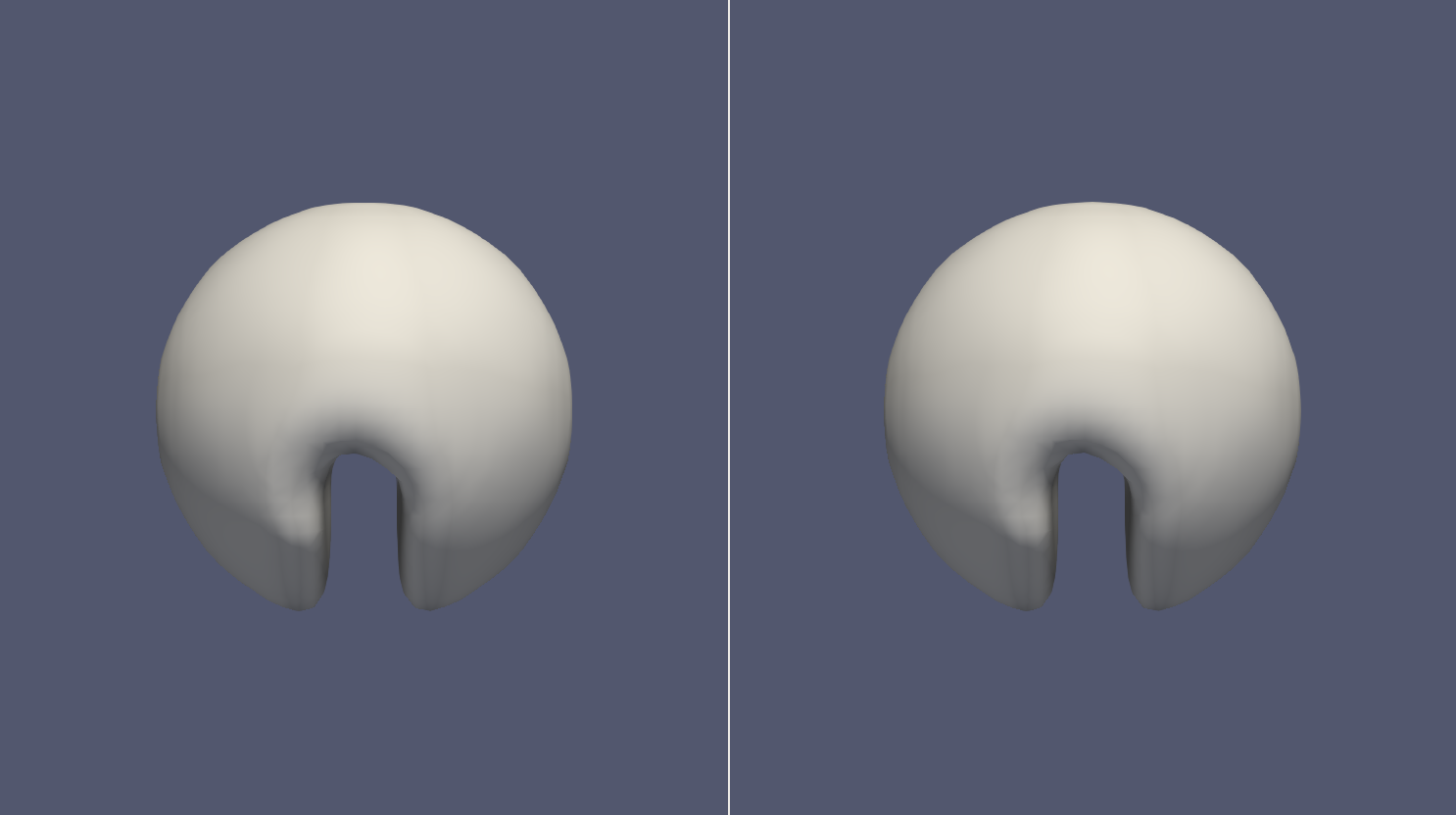}
		\caption{}
	\end{subfigure}	
	\caption{Comparison of the predicted interfaces for the three-dimensional rotating sphere problem with grid number $96^3$. (a) Initial condition; (b) After ten revolutions. (The left-hand side of each figure denotes the solution obtained by \textbf{MPLS} method, and the right-hand side denotes the solution obtained by \textbf{LS} method) }
	\label{disk3d_96}
\end{figure}

\begin{figure}
	
	\centering
	
	\begin{subfigure}[H]{0.6\textwidth}
		\includegraphics[width=\textwidth]{disk3d_ic}
		\caption{}
	\end{subfigure}
	
	\begin{subfigure}[H]{0.6\textwidth}
		\includegraphics[width=\textwidth]{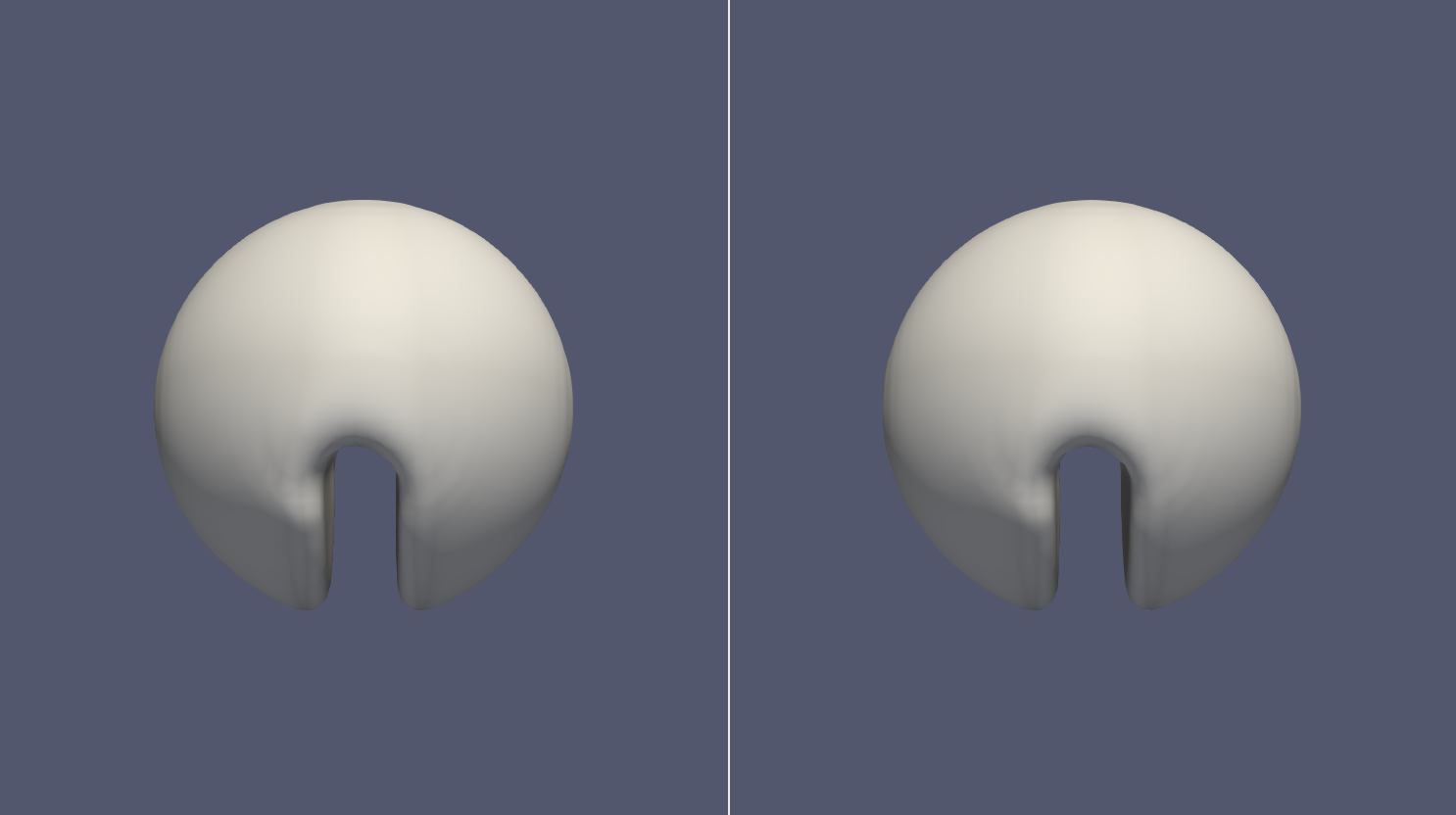}
		\caption{}
	\end{subfigure}	
	\caption{Comparison of the predicted interfaces for the three-dimensional rotating sphere problem with grid number $128^3$. (a) Initial condition; (b) After ten revolutions. (The left-hand side of each figure denotes the solution obtained by \textbf{MPLS} method, and the right-hand side denotes the solution obtained by \textbf{LS} method) }
	\label{disk3d_128}
\end{figure}

\begin{figure}[p]
	\includegraphics[width=\textwidth]{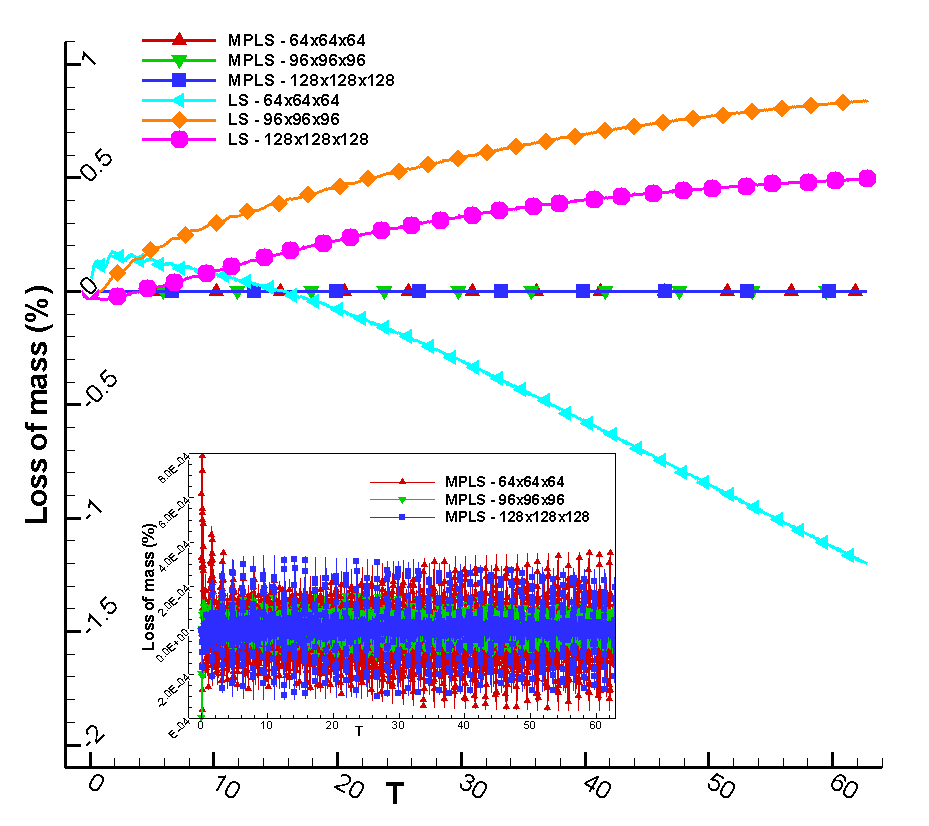}
	\caption{Comparison of the predicted percentages of the loss of mass using different methods with different grid numbers for the three-dimensional rotating sphere problem.}
	\label{disk3d_mass}
\end{figure}

\end{document}